\documentclass[pra,twocolumn,preprintnumbers,amsmath,amssymb,floatfix,longbibliography,superscriptaddress]{revtex4-2}
\usepackage{graphicx}
\usepackage{graphics}
\usepackage{psfrag}
\usepackage{hyperref}
\usepackage{multirow}
\usepackage[sort&compress]{natbib}
\usepackage{amssymb}
\usepackage{amsmath}
\usepackage{amsthm}
\usepackage{bbold}
\usepackage{bbm}
\usepackage{enumerate}
\usepackage{textcomp}
\usepackage{verbatim}
\usepackage{soul}

\newcommand{\vare}{\varepsilon}

\newcommand{\rmi}{{\rm i}}

\newcommand{\vk}{\mathbf{k}}
\newcommand{\vkp}{\mathbf{k}^\prime}
\newcommand{\vq}{\mathbf{q}}
\renewcommand{\vr}{\mathbf{r}}
\newcommand{\vR}{\mathbf{R}}

\DeclareMathOperator{\Tr}{Tr}

\DeclareMathOperator{\diag}{diag}

\renewcommand{\d}{\text{d}}
\DeclareMathOperator{\tr}{tr}
\DeclareMathOperator{\SO}{SO}

\newcommand{\ket}[1]{| #1 \rangle}

\renewcommand{\vec}[1]{\mathbf{#1}}
\newcommand{\hvec}[1]{\hat{\mathbf{#1}}}

\usepackage{xcolor}

\begin{document}

\hypersetup{pdftitle={title}}
\title{Superfluid phase transition of nanoscale-confined helium-3}

\author{Canon Sun}
\email{canon@ualberta.ca}
\affiliation{Department of Physics, University of Alberta, Edmonton, Alberta T6G 2E1, Canada}
\affiliation{Theoretical Physics Institute, University of Alberta, Edmonton, Alberta T6G 2E1, Canada}

\author{Adil Attar}
\affiliation{Department of Physics, University of Alberta, Edmonton, Alberta T6G 2E1, Canada}

\author{Igor Boettcher}
\email{iboettch@ualberta.ca}
\affiliation{Department of Physics, University of Alberta, Edmonton, Alberta T6G 2E1, Canada}
\affiliation{Theoretical Physics Institute, University of Alberta, Edmonton, Alberta T6G 2E1, Canada}

\begin{abstract}
We theoretically investigate the superfluid phase transition of helium-3 under nanoscale confinement of one spatial dimension realized in recent experiments. Instead of the $3\times 3$ complex matrix order parameter found in the three-dimensional system, the quasi two-dimensional superfluid is described by a reduced $3\times 2$ complex matrix. It features a nodal quasiparticle spectrum, regardless of the value of the order parameter.
The origin of the $3 \times 2$ order parameter is first illustrated via the two-particle Cooper problem, where Cooper pairs in the $p_x$ and $p_y$ orbitals are shown to have a lower bound state energy than those in $p_z$ orbitals, hinting at their energetically favorable role at the phase transition.
We then compute the Landau free energy under confinement within the mean-field approximation and show that the critical temperature for condensation of the $3\times 2$ order parameter is larger than for other competing phases. Through exact minimization of the mean-field free energy, we show that mean-field theory predicts precisely two energetically degenerate superfluid orders to emerge at the transition that are not related by symmetry: the A-phase and the planar phase. Beyond the mean-field approximation, we show that strong-coupling corrections favor the A-phase observed in experiment, whereas weak-coupling perturbative renormalization group predicts the planar phase to be stable.
\end{abstract}

\maketitle

Superfluid He-3 is a quantum liquid that exhibits some of the most exotic ground states in condensed matter physics. 
In the absence of a magnetic field, bulk He-3 hosts two stable superfluid phases: the B- and A-phases \cite{vollhardt2013superfluid}. The isotropic, fully-gapped B-phase occupies the vast majority of the pressure-temperature phase diagram and its energetic stability is well understood based on the weak coupling Bardeen--Cooper--Schrieffer (BCS) theory. As a time-reversal symmetric topological superfluid, interest in the B-phase has recently resurfaced as a promising platform to realize Majorana fermions \cite{PhysRevB.78.195125,roy2008topological,PhysRevLett.102.187001,volovik2009topological,volovik2009fermion,PhysRevLett.103.235301,Sato_2017}.
The anistropic, nodal A-phase, on the other hand, is stabilized only at pressures above 21 bar and in a small range of temperatures. Its stability is not captured by weak coupling theory. Instead, spin fluctuations become more pronounced at high pressures and strong-coupling effects need to be incorporated \cite{PhysRevB.13.4745,RevModPhys.47.331}. With two point nodes in the quasiparticle spectrum, the A-phase can be regarded as a superfluid analog of a Weyl semimetal and can host Majorana--Weyl arc states connecting the two nodes \cite{volovik2003universe}.

Recent experiments on He-3 have indicated significant changes to the phase diagram when geometrically confined in the $z$-direction to a quasi-2D setting on scales comparable to the superfluid coherence length \cite{PhysRevLett.124.015301,PhysRevB.104.094520,levitin2013phase,PhysRevLett.111.235304,zhelev2017ab,PhysRevLett.122.085301,PhysRevLett.60.596,PhysRevB.41.11011,PhysRevLett.65.3005}. The most prominent feature is the expansion of the A-phase stability region in the phase diagram, down to zero pressure where weak coupling theory might be applicable \cite{PhysRevB.38.2362,hara1988quasiclassical,nagato1998rough,nagato2000b,PhysRevB.68.064508}. Furthermore, the phase transition from the normal Fermi liquid state to the superfluid phase is always towards the A-phase.
In addition, a third, potentially inhomogeneous phase appears in the phase diagram separating the A- and planar-distorted B-phases, which has been conjectured to be a pair density wave state \cite{PhysRevLett.124.015301,PhysRevLett.128.015301,
PhysRevLett.122.085301,PhysRevLett.98.045301,wiman2016strong,aoyama2016surface,agterberg2020physics,PhysRevB.37.9298,PhysRevB.90.184513}.

In this work, we theoretically elucidate the reason for the dominance of the A-phase of superfluid He-3 in the quasi-2D case under confinement. In the 3D system, the order parameter is a $3\times 3$ complex matrix which transforms under the $s=\ell=1$ representation of the symmetry group $G=\text{U}(1)\times \text{SO}(3)_{\rm S}\times \text{SO}(3)_{\rm L}\times \text{T}$ that captures particle number conservation, spin rotation symmetry, orbital rotation symmetry, and time-reversal symmetry \cite{RevModPhys.59.533}. Our crucial observation is that, under confinement of the $z$-dimension, the superfluid order parameter that condenses at the phase transition is a $3\times 2$ complex matrix instead. It transforms under an irreducible representation (irrep) of $\tilde{G}=\text{U}(1)\times \text{SO}(3)_{\rm S}\times \text{SO}(2)_{\rm L}\times \text{T}$. For temperatures that are much lower than the critical temperature, the usual $3\times 3$ matrix order parameter might become energetically competitive with the $3\times 2$ matrix, similarly spatially inhomogeneous phases, but in this work we will not address resolving this competition and focus on the vicinity of the phase transition instead.

The sole absence of a third column in the order parameter matrix explains several experimentally accessible properties of the superfluid phase transition of He-3 in the quasi-2D case. For one, the reduced order parameter limits the number of possible superfluid phases at the phase transition. In particular, the B-phase is no longer accessible, as observed in Ref. \cite{PhysRevLett.124.015301}. Furthermore, no other fully gapped phase is accessible, regardless of the precise values of the quartic coefficients in the Ginzburg--Landau free energy functional, since the algebraic structure of the $3\times 2$ matrix order parameter immediately implies nodal points in the quasiparticle excitation spectrum.

To illustrate the emergence of the $3\times 2$ order parameter at the phase transition, we first study the Cooper instability of the Fermi surface under confinement in Sec. \ref{sec: Cooper}. We show that the degeneracy between the three $p$-wave Cooper pair bound states is lifted under confinement, and Cooper pairs in the $p_x$ and $p_y$ orbitals have a lower energy, hence are energetically favored to undergo condensation. In Sec. \ref{sec: MF}, we confirm this perspective within the framework of mean-field theory. We show that the $3\times 3$ order parameter splits into two sub-order parameters with different transition temperatures: a $3\times 2$ matrix and a $3\times 1$ matrix. We find that the $3\times 2$ matrix has a higher transition temperature and is thus the only relevant order parameter at the superfluid transition.

A fascinating problem, inherent to symmetry breaking described by matrix order parameters in general, concerns the question of which particular matrix is chosen during spontaneous symmetry breaking. In the present case, the Ginzburg--Landau free energy functional derived from mean-field theory features many energetically degenerate ground states. For instance, mean-field theory does not dictate whether time-reversal symmetry is broken or not at the transition. This degeneracy of ground states at the mean-field level is common in similar systems such as bulk He-3 \cite{vollhardt2013superfluid}, d-wave superconductivity in 3D \cite{PhysRevA.9.868,PhysRevB.100.104503}, triple-point fermions \cite{PhysRevB.104.L180507}, or complex tensor order \cite{PhysRevLett.120.057002,PhysRevB.101.184503}. Remarkably, we find that in the present case, the many degenerate ground states actually only consists of two physically inequivalent configurations, the A-phase and the planar phase, whereas all other minima are related to these two via a symmetry transformation from $\tilde{G}$.

To resolve the question of whether the A-phase or the planar phase is energetically favored, we apply two theoretical approaches. First, we use the best available strong-coupling corrections for the Landau free energy from the 3D system and confirm that the A-phase is favored at the transition for all pressures. Second, we use weak-coupling perturbative renormalization group (RG) to incorporate order parameter fluctuations in the confined system, which predicts the planar phase to be favorable instead. This mismatch with observations indicates that even low pressures do not correspond to a weak coupling, but we also speculate on general issues of the perturbative RG to resolve second-order phase transitions of matrix order parameters. The lifting of the mean-field degeneracy due to fluctuations in the RG is unlike the case discussed in Ref. \cite{PhysRevB.104.L180507}, where the degeneracy is protected by an enlarged symmetry. Experimentally, measuring the absence or presence of time-reversal symmetry at the transition could determine the condensation of the A-phase or planar phase, respectively.

\section{Physical picture: The Cooper problem}\label{sec: Cooper}

To illuminate the effect of confinement on the superfluid state, we first examine the associated Cooper problem. This amounts to solving the two-particle Schr\"{o}dinger equation in the presence of a static Fermi surface (FS). The formation of a bound state, the Cooper pair, and its functional dependence on the system parameters is often a good approximation of the superfluid instability found within mean-field theory for the actual many-body system, such as in spin- or mass-imbalanced ultracold Fermi gases \cite{PhysRevA.89.063609}.

In the original Cooper problem, two atoms in 3D that experience a two-dimensional FS are interacting attractively, which leads to the formation of a shallow bound state. The atoms are described by the two-particle Schr\"{o}dinger equation
\begin{align}
    \label{coop1} \left[-\frac{\hbar^2}{4m}\nabla_{\vR}^2-\frac{\hbar^2}{m}\nabla_{\vr}^2-2\mu-V(r)\right]\psi(\vR,\vr)=E\psi(\vR,\vr),
\end{align}
where $\vR=(X,Y,Z)^T$ and $\vr=(x,y,z)^T$ are the center-of-mass and relative coordinates of the two atoms, respectively, $\mu$ is the chemical potential, and $V(r)$ the interaction potential. Here $\psi(\vR,\vr)$ is the orbital part of the two-particle wavefunction. The spin part of the wavefunction is assumed to be a triplet state and has been factored out. To model confinement to quasi-2D, place the atoms in a box with length $L_i$ in the $i=x,y,z$ direction. The system is confined in the $z$ direction in that $L_z$ is comparable to the superfluid coherence length $\xi=70\ \text{nm}$ of the bulk 3D system, whereas $L_{x},L_y\gg \xi$. The Schr\"{o}dinger equation is then solved with Dirichlet boundary conditions for the center-of-mass coordinate $Z$, i.e. $\psi(X,Y,Z=0,\vr)=\psi(X,Y,Z=L_z,\vr)=0$, and periodic boundary conditions for the other coordinates.

In order to solve Eq. (\ref{coop1}), it is convenient to expand the wavefunction in basis functions that are compatible with the boundary conditions according to $\psi(\vR,\vr)=\sum_{\vq,\vk} \sin (q_z Z)e^{i(q_x X+q_y Y)}e^{i\vk\cdot \vr}\psi_\vq(\vk)/\mathcal{V}^2$.  Here $\vq=(q_x,q_y,q_z)^T$ and $\vk=(k_x,k_y,k_z)^T$ denote the center-of-mass and relative wavevectors, respectively, and $\mathcal{V}=L_xL_yL_z$ is the system volume. The wavevector $q_z$ has a different quantization condition arising from the different boundary condition. We have $q_z=\pi n/L_z$ whereas $q_{x,y}=2\pi m_{x,y}/L_{x,y}$ and $k_{x,y,z}=2\pi l_{x,y,z}/L_{x,y,z}$, with $n=1,2,\dots$ and $m_{x,y},l_{x,y,z}\in \mathbb{Z}$. In this basis, the Cooper problem reads
\begin{equation}
    \left(\xi_{\vk_+}+\xi_{\vk_-}\right)\psi_{\vq}(\vk)-\frac{1}{\mathcal{V}}\sum_{\vkp}V(\vk-\vkp)\psi_\vq(\vkp)=E\psi_{\vq}(\vk),
\end{equation}
where  $\vk_{\pm}=\vk\pm \vq/2$, $\xi_\vk=\hbar^2k^2/(2m)-\mu$ is the single-particle energy dispersion, and $V(\vk)=\int \d^3 r e^{-i\vk\cdot\vr}V(r)$ is the interaction matrix element.  In He-3, the dominant angular momentum channel is $l=1$, and we choose $V(\vk-\vkp) =4\pi g \sum_{m=-1}^1Y_{1m}(\hat{\vk})Y_{1m}^*(\hat{\vkp})$,
where $g>0$ is the coupling constant and $Y_{lm}(\hat{\vk})=Y_{lm}(\theta_{\textbf{k}},\phi_{\textbf{k}})$ is the spherical harmonic of degree $l$ and order $m$. 
The corresponding Hamiltonian is time-reversal symmetric and, importantly, conserves the center-of-mass momentum $\vq$. Generally, a large $\vq$ will increase the energy of the bound state. Thus, we take $\vq$ to be the smallest admissible wavevector $\vq=(0,0,\pi/L_z)$.
This breaks the three-dimensional spatial rotational symmetry, leaving only SO(2) rotation symmetry about the $z$-axis. This contrasts with an unconfined system in which $q_z$ can be arbitrarily small.

Crucially, because of the Pauli exclusion principle, the Cooper pair must live on top of the FS. This amounts to imposing the constraint $\psi_{\vq}(\vk)=0$ whenever $\xi_{\vk_+}$ or $\xi_{\vk_-}$ is negative. Thus the wavefunction can be assumed to take the form $\psi_{\vq}(\vk)=\Theta(\xi_{\vk_+})\Theta(\xi_{\vk_-}) \phi_{\vq}(\vk)$, where $\Theta(x)$ is the Heaviside step function. As the Hamiltonian is SO(2) symmetric, it is convenient to decompose the Schr\"{o}dinger equation into SO(2) angular momentum channels. To this end, we express the wavefunction and potential in SO(2) partial waves according to $\phi_{\vq}(\vk)=\sum_{m\in\mathbb{Z}} \phi_{\vq,m}(k,\theta_\vk)e^{im\phi_\vk}$ and $V(\vk,\vkp)=4\pi g\sum_{m=-1}^1 Y_{1m}(\theta_\vk,0)Y_{1m}(\theta_{\vkp},0)e^{im(\phi_\vk-\phi_{\vkp})}$. 
For the angular momentum channels $m=0,\pm 1$, the Cooper problem becomes
\begin{equation}\label{eq: Cooper pair gap equation}
	\frac{1}{4\pi g}=\int \frac{ \d k \d\theta_\vk}{(2\pi)^2} k^2 \sin\theta_\vk\frac{\Theta(\xi_{\vk_+})\Theta(\xi_{\vk_-})}{\xi_{\vk_+}+\xi_{\vk_-}-E}  Y_{1m}^2(\theta_\vk,0).
\end{equation}
A negative energy solution to Eq. \eqref{eq: Cooper pair gap equation} signifies a Cooper pair bound state. Note that, because $Y_{1,1}^2(\theta_\vk,0)=Y_{1,-1}^2(\theta_\vk,0)$, the $m=+1$ and $m=-1$ states have the same energy, which is a consequence of time-reversal symmetry.

\begin{figure}[t!]
 \centering
   \includegraphics[width=0.9\columnwidth]{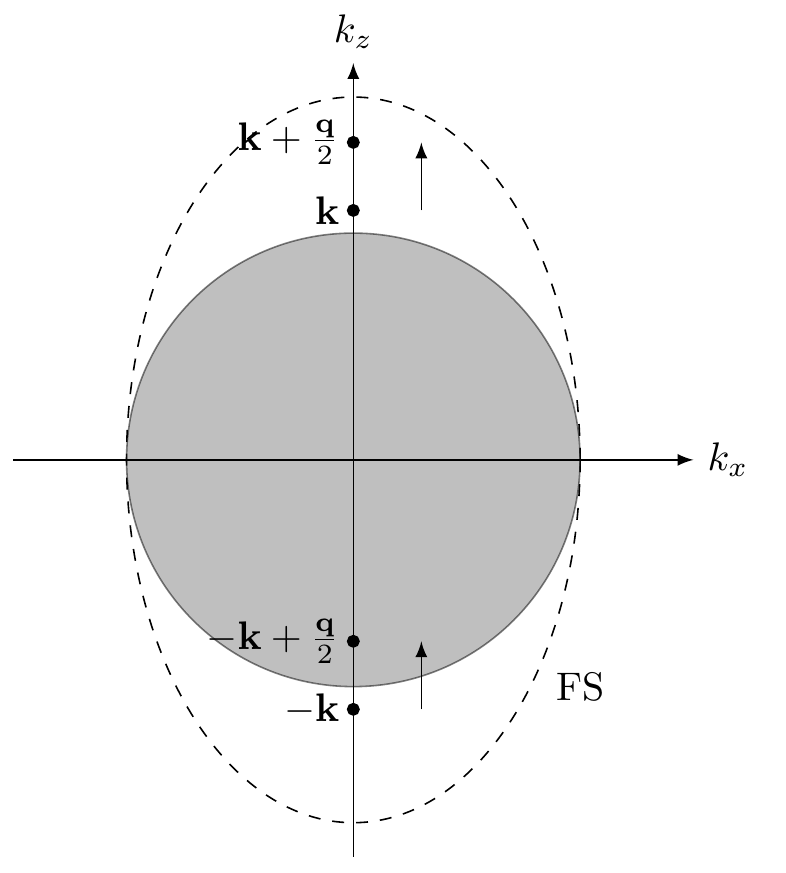}
 \caption{\textbf{Cooper pairing under confinement.} In this schematic, helium atoms are depicted as black circles above the Fermi surface (gray). When $\vq=0$, a Cooper pair can be formed as long as the atoms are above the FS. When $\vq\neq 0$, however, a Cooper pair cannot be formed within the dashed line because at least one atom will be in the Fermi sea. }
 \label{fig: FS as an obstacle}
\end{figure}

The center-of-mass wavevector $\vq$ diminishes the region in momentum space where the helium atoms can interact attractively. Making the change of variable $k\mapsto x=\hbar^2k^2/2m-\mu$, Eq. \eqref{eq: Cooper pair gap equation} becomes
\begin{align}\label{eq: Cooper pair gap equation 2}
\begin{split}
    \frac{1}{4\pi N(0)g}=&\int_0^\Lambda \d x \int_0^\pi \d\theta_\vk\frac{\sin\theta_\vk}{2} \\&\frac{\Theta(x-\hbar v_{\rm F} q|\cos\theta_\vk|/2)}{2x-E^\prime} Y_{1m}^2(\theta_\vk,0),
    \end{split}
\end{align}
where $N(0)=(2m/\hbar^2)^{3/2}\sqrt{\mu}/(4\pi^2)$ is the density of states per spin at the Fermi level, $E^\prime=E-2\epsilon_{\vq/2}$, $\epsilon_\vk=\hbar^2k^2/(2m)$, and $\Lambda$ the energy cutoff. For $q\neq 0$, in contrast to the unconfined case,  the step function increases the lower limit of the $x$ integral, reducing the domain of integration. Physically, a non-zero center-of-mass momentum presents an obstacle to forming Cooper pairs in the vicinity of the FS. Take, for example, a pair of atoms with wavevectors $\pm (0,0,k_z)$, where $k_{\rm F}<k_z<k_{\rm F}+q/2$, as shown in Fig. \ref{fig: FS as an obstacle}. Now suppose they acquire a non-zero $\vq$, i.e. their wavevectors become $\pm (0,0,k_z+q/2)$. Then one of the atoms enter the Fermi sea, which is prohibited by the exclusion principle. Thus, when $\vq\neq 0$, the region where atoms cannot pair extends beyond the Fermi sea.

Performing the integrals in Eq. \eqref{eq: Cooper pair gap equation 2}, when $\hbar v_{\rm F} q< |E^\prime|$, the energy levels in the $m=0$ angular momentum channel satisfy the transcendental equation
\begin{align}
    \frac{1}{ N(0)g}=\frac{1}{2}\ln\left(\frac{E^\prime-2\Lambda}{E^\prime} \right)+\frac{3y}{8}{}_3F_2(1,1,4;2,5;y),
\end{align}
whereas for $m=1$ we obtain
\begin{align}
    \frac{1}{ N(0)g}=\frac{1}{2}\ln\left(\frac{E^\prime-2\Lambda}{E^\prime} \right)+\frac{3y}{16} {}_4F_3(1,1,2,4;2,3,5;y).
\end{align}
Here ${}_pF_q(a_1,\dots,a_p;b_1,\dots,b_q;y)$ is the generalized hypergeometric function and $y=\hbar v_{\rm F} q/E^\prime$. In the experimentally accessible regime, $\hbar v_{\rm F} q\ll |E|$, the bound state energy of a Cooper in the angular momentum channel $m$ is
\begin{equation}
    E_m\simeq -2\Lambda e^{-2/(N(0)g)}+
    \begin{cases}
        3\hbar v_{\rm F} q/4&m=0\\
        3\hbar v_{\rm F} q/8&m=\pm 1
    \end{cases}.
\end{equation}
Importantly, the $m=\pm 1$ bound states have a lower energy than $m=0$. This result has a simple interpretation. When $\vq=0$, the angular parts of the Cooper pair wavefunctions are precisely the $l=1$ spherical harmonics, $Y_{1m}(\theta_\vk,\phi_\vk)$. The center-of-mass wavevector reduces the region in momentum space where the atoms can pair. This reduction is not isotropic, but most prominent near the poles. Thus, atoms in the $Y_{10}(\theta_\vk,\phi_\vk)$ state, whose probability density is more concentrated at the poles compared to the $Y_{1,\pm 1}(\theta_\vk,\phi_\vk)$ states, have a greater reduction in momentum space where they can interact attractively, and therefore a higher energy.

\section{Mean-field theory}\label{sec: MF}
The above simple analysis illustrates that, under quasi-2D confinement, the Fermi surface is unstable to the formation of Cooper pairs in the $p_x$ and $p_y$ orbitals. Thus we expect only atoms in these two orbitals to condense at the superfluid phase transition. In this section, we provide a many-body justification of this picture within the mean-field approximation.

\subsection{Free energy}
As illustrated above, the crucial element under confinement is that the Cooper pairs acquire a non-zero center-of-mass wavevector $\vq$. To incorporate this into the mean-field analysis, we consider the imaginary-time action 
\begin{equation}
    F[\psi]=\frac{1}{2}\left(F_\vq[\psi]+F_{-\vq}[\psi]\right),
\end{equation}
where
\begin{align}
&F_\vq[\psi]=\int_0^\beta\d \tau \bigg[\sum_{\vk s}\bar{\psi}_{\vk s}(\partial_\tau-\xi_{\vk} )\psi_{\vk s}\\
&-\frac{1}{2\mathcal{V}}\sum_{ \substack{\vk\vkp \\ s_1s_2 s_3s_4}
  }\bar{\psi}_{\vk_+s_1}\bar{\psi}_{-\vk_- s_2}V_{s_1s_2s_3s_4}(\vk,\vkp)\psi_{-\vkp_-s_3}\psi_{\vkp_+s_4}\bigg].\nonumber
\end{align}
Here $\psi_{\vk s}$ is a 
Grassmann field with wavevector $\vk$ and spin $s=\uparrow,\downarrow$, $\beta=1/(k_{\rm B}T)$ is the inverse temperature, and $\vk_{\pm}=\vk\pm \vq/2$ as before.  The interaction matrix element remains SO(3) symmetric under confinement and takes the p-wave form
\begin{align}
 \nonumber V_{s_1s_2s_3s_4}(\vk,\vkp)={}& 2\pi g \sum_{\mu=1}^3 \sum_{m=-1,0,1} Y_{1m}(\vk) Y^*_{1m}(\vkp)\\
&\times ( \sigma_\mu i\sigma_y)_{s_1s_2}( -i\sigma_y \sigma_\mu)_{s_3s_4}.
\end{align}
We denote the Pauli spin matrices by $\sigma_\mu$, with index $\mu=1,2,3=x,y,z$. The wavevector sums over $\vk$ and $\vkp$ are restricted to a momentum shell of thickness $\Lambda$ around the Fermi surface. The action $S_\vq[\psi]$ describes He-3 atoms interacting attractively in the Cooper channel, except that the Cooper pairs have a conserved center-of-mass momentum $\vq$, and $S[\psi]$ consists of contributions from the two Fourier modes $\vq$ and $-\vq$. The inclusion of both Fourier modes is necessary in order to satisfy the boundary conditions.

In anticipation of a $p$-wave superfluid transition, we introduce the $3\times 3$ complex matrix order parameter 
\begin{equation}
    A_{\mu j}(\vq)=\frac{g}{2\mathcal{V}}\sum_{\vk, s_1s_2 }\langle c_{-\vk_-s_1}c_{\vk_+s_2} \rangle (i\sigma_y\sigma_\mu )_{s_1s_2}\frac{\sqrt{3}k_j}{k_{\rm F}}.
\end{equation}
A nonvanishing $A_{\mu j}(\vq)$ signifies the condensation of Cooper pairs in the $p_j$ orbital with zero spin projection in the $\mu$ direction. In order to satisfy the boundary conditions, the order parameter must satisfy $A_{\mu j}(-\vq)=-A_{\mu j}(\vq)$.
 The superfluid transition can be captured with a bosonic theory for the order parameter $A_{\mu i}(\vq)$. Performing a Hubbard--Stratonovich transformation in the Cooper channel and integrating out the fermions, we obtain the trace-log effective action
 \begin{equation}\label{eq: MF exact action}
F[A]=\frac{\beta \mathcal{V}}{g}\sum_{\mu ,j}A^*_{\mu j}(\vq)A_{\mu j}(\vq)-\frac{1}{2}\sum_{k}\tr\ln \mathcal{G}^{-1}(\vq,k).
\end{equation}
Herein, $\mathcal{G}^{-1}(\vq,k)=ik_n-H(\vq,\vk)$ is the inverse Nambu--Gorkov Green function and
\begin{equation}
     H(\vq,\vk)=\begin{bmatrix}
\xi_{\vk_+}\mathbb{1}_2&\Delta(\vq,\vk)\\
\Delta^\dagger(\vq,\vk)&-\xi_{-\vk_-}\mathbb{1}_2
    \end{bmatrix}
\end{equation}
is the Bogoliubov-de Gennes (BdG) Hamiltonian. We denote the $2\times 2$ unit matrix by $\mathbb{1}_2$ and the gap matrix by 
\begin{equation}\label{eq: gap}
    \Delta_{s_1s_2}(\vq,\vk)=\sum_{\mu, j}A_{\mu j}(\vq) (\sigma_\mu i\sigma_y)_{s_1s_2}\frac{\sqrt{3} k_j}{k_{\rm F}}.
\end{equation}
Bosonic Matsubara frequencies are given by $k_n=2\pi n/\beta$ with $n\in\mathbb{Z}$ and we employ the short-hand notation $k=(ik_n,\vk)^T$. 

The nature of the phase transition can be understood from the Ginzburg--Landau free energy expanded to quartic order in $A_{\mu j}$,
\begin{equation}\label{eq: G-L}
    F[A]= F^{(2)}[A]+F^{(4)}[A]+\dots,
\end{equation}
where $F^{(2)}$ ($F^{(4)}$) is quadratic (quartic) in the order parameter and $+\dots$ denotes higher-order terms. 
In the following, we summarize the result of the expansion. The detailed calculation can be be found in App. \ref{app: MF coefficients}. The quadratic term in the Ginzburg--Landau free energy determines the form of the order parameter and transition temperature. 
In the mean-field regime we find
\begin{equation}\label{eq: S2}
    F^{(2)}[A]=\alpha_{xy}(\vq)\Tr \tilde{A}^\dagger \tilde{A}+\alpha_{z}(\vq)\Tr A_{\rm z}^\dagger A_{\rm z}.
\end{equation}
In contrast to the unconfined case, the $3\times 3$ matrix order parameter $A_{\mu j}$ is split into a $3\times 2$ matrix $\tilde{A}$  and a $3\times 1$ matrix $A_{\rm z}$ because of the broken SO(3) symmetry. The two order parameters are defined as 
\begin{align}
\tilde{A}&=\begin{pmatrix} A_{xx} & A_{xy} \\ A_{yx} & A_{yy} \\ A_{zx} & A_{zy} \end{pmatrix},& A_{\rm z} &= \begin{pmatrix} A_{xz} \\ A_{yz} \\ A_{zz} \end{pmatrix}.
\end{align}
Thus $\tilde{A}$ comprises the first two columns of $A$, $A_{\rm z}$ the third column, and we can write 
\begin{align}
A=(\tilde{A}\ A_{\rm z})
\end{align}
in matrix notation. Physically, $\tilde{A}\neq 0$ signals the condensation of Cooper pairs in the $p_x$ and $p_y$ orbitals, and $A_z\neq 0$ in the $p_z$ orbital. 

The mean-field quadratic coefficients in the confined quasi-2D geometry are 
\begin{align}
\alpha_{xy}(\vq)&\simeq \beta \mathcal{V}\left[\alpha(0) +\frac{7\zeta (3)}{20\pi^2} \frac{N(0)\mu}{(k_{\rm B}T)^2}\epsilon_\vq\right]\label{eq: alphaxy}\\
\alpha_z(\vq)&\simeq \beta \mathcal{V}\left[\alpha(0) +\frac{21\zeta (3)}{20\pi^2} \frac{N(0)\mu}{(k_{\rm B}T)^2}\epsilon_\vq\right]\label{eq: alphaz},
\end{align}
where $
 \epsilon_{\textbf{q}} =\hbar^2 q^2/(2m)$,
$\alpha(0)=1/g-N(0)\ln \left(2e^\gamma\Lambda/(\pi k_{\rm B} T) \right)$, $\zeta(z)$ is the Riemann zeta function, and $\gamma\simeq 0.5772$ is the Euler-Mascheroni constant. The Cooper logarithmic divergence persists under confinement, which guarantees a superfluid transition at a sufficiently low temperature. Importantly, $\alpha_{xy}(\vq)\leq \alpha_z(\vq)$. This implies that the order parameter $\tilde{A}$ has a higher transition temperature than $A_{\rm z}$ for $q\neq 0$. Consequently, under confinement, the phase transition from the Fermi liquid phase is captured by the $3\times 2$ order parameter $\tilde{A}$ as opposed to the full $3\times 3$ matrix $A$ in the unconfined case.

The specific reduction of the order parameter under confinement is a result of the explicit symmetry breaking from $\text{SO}(3)_{\rm L}$ to $\text{SO}(2)_{\rm L}$ in conjunction with time-reversal symmetry. In the 3D system, the order parameter transforms under an irrep of $G$, which is labelled by the spin and orbital angular momentum quantum numbers $s$ and $l$, and is a $(2s+1)\times (2l+1)$ matrix, which $s=l=1$, and so the order parameter is a $3\times 3$ matrix. In the confined quasi-2D system, the introduction of $\vq$ breaks the symmetry group from $G$ to $\tilde{G}$. 
While the spin part of the order parameter remains unchanged, under the restriction to $\text{SO}(2)_{\rm L}$, the irreps are labeled by the azimuthal quantum number $m$. As a result, the $p$-wave order parameter splits into three components with orbital basis functions $Y_{1m}(\hat{\vk})$, where $m=0,\pm 1$. However, time-reversal maps $Y_{1,1}(\vk)$ and $Y_{1,-1}(\vk)$ onto each other, stiching the irreps $m=+1$ and $m=-1$ together into a single irrep. Thus, the $p$-wave order parameter separates into two: one with orbital basis functions $Y_{11}(\vk)$ and $Y_{1,-1}(\vk)$, or equivalently, $k_x$ and $k_y$, which corresponds to $\tilde{A}$, and the other with $Y_{10}(\vk)\sim k_z$, corresponding to $A_{\rm z}$. Note that time-reversal symmetry does not play a role in the classification of irreps in 3D as all representations of $\text{SO}(3)_{\rm L}$ are real or pseudoreal, whereas it does under confinement since the irreps of $\text{SO}(2)_{\rm L}$, except for $m=0$, are all complex.

An important consequence of the reduced order parameter is that the superfluid phase under confinement is necessarily nodal. Indeed, as the components $(A_{\rm z})_\mu=A_{\mu z}$ of the order parameter do not condense and are thus zero, the pairing matrix $\Delta(\vq,\vk)$ in Eq. \eqref{eq: gap} vanishes at the poles of the FS given by the momenta
\begin{align}
\vk_0=\pm (0,0,k_{\rm F})^T.
\end{align}
This is the case \emph{regardless} of the components of $\tilde{A}$. Consequently, the quasiparticle spectrum in the confined case (the spectrum of $H(\vq,\vk)$) is always gapless in the vicinity of the phase transition. This is one of the key findings of the present analysis.

The actual ground state and ensuing order parameter $\tilde{A}$ at a second-order phase transition is determined by the quartic coefficients in the Ginzburg--Landau free energy. The most general contribution to the free energy quartic in $\tilde{A}$ compatible with the symmetry is
\begin{equation}
    F^{(4)}[\tilde{A}]=\sum_{a=1}^5 \beta_a(\vq) \tilde{I}_a(\tilde{A}),
\end{equation}
with
\begin{align}
    \tilde{I}_1(\tilde{A})&=|\tr (\tilde{A}\tilde{A}^T)|^2,\\
    \tilde{I}_2(\tilde{A})&=[\tr (\tilde{A} \tilde{A}^\dagger)]^2,\\
    \tilde{I}_3(\tilde{A})&=\tr[(\tilde{A}\tilde{A}^T)(\tilde{A}\tilde{A}^T)^*],\\
 \tilde{I}_4(\tilde{A})&=\tr[(\tilde{A}\tilde{A}^\dagger)^2],\\
\tilde{I}_5(\tilde{A})&=\tr[(\tilde{A}\tilde{A}^\dagger)(\tilde{A}\tilde{A}^\dagger)^*].
\end{align}
These are the usual five quartic invariants known from the study of He-3 in 3D, evaluated here, however, for the $3\times 2$ order parameter. The quartic coupling constants calculated from mean-field theory satisfy the relation
\begin{equation}
-2\beta_1(\vq)=\beta_2(\vq)=\beta_3(\vq)=\beta_4(\vq)=-\beta_5(\vq),
\end{equation}
with
\begin{align}\label{eq: beta2}
    \beta_2(\vq)&=\frac{21\zeta(3)}{40\pi^2}\frac{N(0)}{(k_{\rm B}T)^2}-\frac{279\zeta(5)}{560\pi^4}\frac{N(0)\mu}{(k_{\rm B}T)^4}\epsilon_\vq.
\end{align}
Remarkably, these ratios in the quasi-2D system with $q\neq 0$ are identical to those found for the 3D system with $q=0$. The details of the calculation are presented in App. \ref{app: MF coefficients}.

\subsection{Minimizing the free energy}
Remarkably, the mean-field Landau free energy for the reduced order parameter can be minimized exactly. Let us write $\tilde{A}= \Delta_0 \tilde{A}^\prime$, where $\Delta_0$ is a complex number and $\tilde{A}^\prime$ a $3\times 2$ complex matrix normalized such that 
\begin{equation}\label{eq: A normalization}
\tr \tilde{A}^{\prime\dagger} \tilde{A}^\prime=1.
\end{equation}
Here $\Delta_0$ characterizes the magnitude of the order parameter and $\tilde{A}^\prime$ the internal structure of the Cooper pair. In this parameterization, the free energy becomes
\begin{align}
F[\Delta_0,\tilde{A}^\prime]=\alpha_{xy} |\Delta_0|^2  +\sum_{a=1}^5 \beta_a  \tilde{I}_a(\tilde{A}^\prime)|\Delta_0|^4.
\end{align}
Minimizing the free energy with respect to $\Delta_0$ for $\alpha_{xy}<0$, the minimum occurs at $|\Delta_0|^2=-\alpha_{xy}/[2\sum_{a}\beta_a \tilde{I}_a(\tilde{A}^\prime)]$. At this value of $\Delta_0$, the free energy is
\begin{equation}
	F_{\text{min}}[\tilde{A}^\prime]=-\frac{\alpha}{4\sum_{a=1}^5 \beta_a \tilde{I}_a(\tilde{A}^\prime)}.
\end{equation}
Thus, the free energy is minimized when the quartic contribution to the free energy,
\begin{equation}
	F^{(4)}[\tilde{A}^\prime]=\sum_{a=1}^5 \beta_a \tilde{I}_a(\tilde{A}^\prime),
\end{equation}
is minimized subject to the condition Eq. \eqref{eq: A normalization}. 

We now proceed to minimize the quartic free energy. Here we only outline the procedure and summarize the results. The detailed calculation is presented in App. \ref{app: energy minimization}. First, we establish a lower bound for the quartic free energy. If an order parameter configuration saturates the bound, then it is necessarily a ground state.
To this end, we reorganize the quartic free energy as
\begin{equation}
    F^{(4)}[\tilde{A}^\prime]=\beta_2\left[1+\left(\tilde{I}_3-\frac{\tilde{I}_1}{2}\right)+\left(\tilde{I}_4-\tilde{I}_5\right)\right].
\end{equation}
Here we have used that $\tilde{I}_2(\tilde{A}^\prime)=1$ by the normalization condition Eq. \eqref{eq: A normalization}. The motivation for such groupings of invariants originates from the following inequalities:
\begin{align}
    \tilde{I}_3-\frac{\tilde{I}_1}{2}&\geq 0\label{eq: inequality 31} \\
    \tilde{I}_4-\tilde{I}_5&\geq 0.\label{eq: inequality 45}
\end{align}
As $\beta_2>0$, if a state saturates the bounds in both inequalities, then it is a ground state. Based on this, we find only two ground states not related by symmetry transformations. The first is the time-reversal symmetry breaking A-phase, with order parameter
\begin{equation}
    \tilde{A}^\prime=\frac{1}{\sqrt{2}}\begin{pmatrix}
        0&0\\
        0&0\\
        1&i
    \end{pmatrix}.
\end{equation}
The other is the time-reversal symmetry preserving planar phase,
\begin{equation}
    \tilde{A}^\prime=\frac{1}{\sqrt{2}}\begin{pmatrix}
        1&0\\
        0&1\\
        0&0
    \end{pmatrix}.
\end{equation}
These two phases define two symmetry-inequivalent classes of ground states.

\subsection{Strong-coupling corrections}

\begin{figure}
    \centering
   \includegraphics[width=0.98\columnwidth]{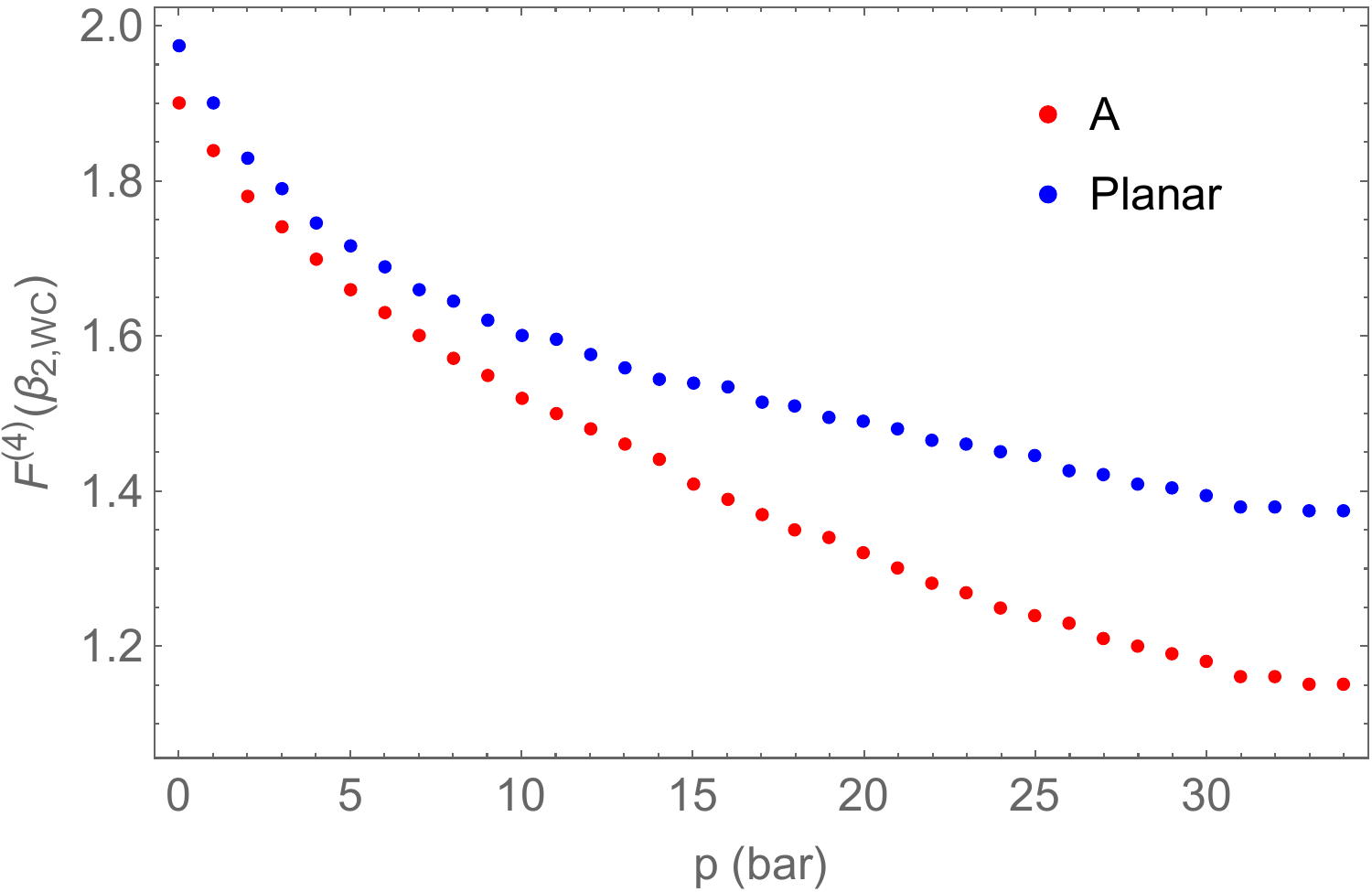}
    \caption{\textbf{Strong-coupling corrections.} We display the quartic free energy of the A-phase (red dots) and planar phases (blue dots) from strong-coupling corrections at the critical temperature vs. pressure. We observe the A-phase to have a lower free energy and hence to be the stable superfluid ground state. The quartic free energy is normalized to the weak-coupling value for $\beta_2$.} 
    \label{fig: strong coupling}
\end{figure}

To study the relative stability of the two classes of ground states, we compare their quartic free energies when strong coupling corrections beyond the mean-field approximation are incorporated. In Ref. \cite{PhysRevB.75.174503}, the $\beta_a$-coefficients of $F[A]$ at various pressures were obtained experimentally by measurements of, for example, the nuclear magnetic resonance $g$-shift or the specific heat jump. These experiments were performed on a 3D system. However, as our above calculation shows, the effects of confinement on the $\beta_a$-coefficients are small, and thus it is justified to take the bulk values.  Using these strong-coupling coefficients, we compute the quartic free energy of the A- and planar phases at the critical temperature and various pressures, as shown in Fig. \ref{fig: strong coupling}. The degeneracy at weak coupling is lifted upon the inclusion of strong-coupling corrections. We find that the A-phase always has a lower energy than the planar phase.

\section{Perturbative renormalization group analysis}\label{sec: RG}

Mean-field theory predicts a second-order phase transition towards superfluidity at a critical temperature $T_{\rm c}$ with $3\times 2$ complex matrix order parameter $\tilde{A}_{\mu i}$. Since the order parameter vanishes directly at the phase transition, its fluctuations can be substantial. These fluctuation effects are not captured by mean-field theory, but can be incorporated  with the RG \cite{ZinnJustinBook,herbutbook}. In the following, we discuss the predictions from perturbative RG for the phase transition in nanoscale-confined He-3, with the full calculation presented in App. \ref{AppBRG}.

The bosonic theory for the order parameter field $\tilde{A}(\tau,\textbf{R})$ after integrating out the fermions is given by the mean-field Ginzburg--Landau free energy, to quartic order in the field,
\begin{align}
\nonumber F_{\rm MF}[\tilde{A}] ={}& \int_0^{1/T}\mbox{d}\tau \int \mbox{d}^2R_\perp \int_0^{L_z}\mbox{d}Z  \\
\nonumber &\Biggl[\mathcal{Z}_{\rm MF} \tilde{A}_{\mu_i}^*\partial_\tau \tilde{A}_{\mu i}+K_{\rm MF}\nabla \tilde{A}_{\mu i}^*\cdot \nabla \tilde{A}_{\mu i}\\
&+ K_{\rm MF}(\gamma_{\rm MF}-1)(\partial_i \tilde{A}_{\mu i}^*)(\partial_j \tilde{A}_{\mu j}) \\
\label{bos1} &+ \alpha_{\rm MF}\ \mbox{tr}(\tilde{A}^\dagger \tilde{A})+\sum_{a=1}^5 \beta_{a,\rm MF} \tilde{I}_a\Biggr],
\end{align}
Here $\tau$ is imaginary time and $\textbf{R}=(\textbf{R}_\perp,Z)=(X,Y,Z)$ is the center of mass coordinate of the Cooper pair. The variable $\textbf{R}$ is conjugate to the momentum $\textbf{q}$ from Sec. \ref{sec: MF}.  The coordinates in the unconfined directions, $\textbf{R}_\perp=(X,Y)$, are assumed to be infinitely extended. In contrast, the z-direction is confined to the interval $Z\in[0,L_z]$ with Dirichlet boundary conditions for $\tilde{A}$ at the end points. In Eq. (\ref{bos1}), repeated indices are summed, with $\mu =1,2,3,\ i=1,2$. The mean-field coefficients $\mathcal{Z}_{\rm MF}, K_{\rm MF}, \alpha_{\rm MF},\beta_{a,\rm MF}$ depend on the thermodynamic variables $(T,\mu)$ or $(T,P)$, whereas $\gamma_{\rm MF}=3$. There are two kinetic terms: the first, with $\nabla=(\partial_X,\partial_Y,\partial_Z)$, contains derivatives in all three spatial directions, whereas the second one proportional to $\gamma-1$ only depends on $\partial_X$ and $\partial_Y$.

Finite temperature is incorporated in Eq. (\ref{bos1}) through the imaginary time $\tau$ such that the field $\tilde{A}(\tau,\textbf{R})$ is periodic in $\tau$ with period $1/T$. As a result, temporal fluctuations of $\tilde{A}$ have quantized bosonic Matsubara frequencies $2\pi \nu T$ with $\nu\in\mathbb{Z}$. These are energetically unfavorable unless $\nu=0$, so that we can ignore the bosonic modes with $\nu \neq 0$. This, on the other hand, implies that $\tilde{A}(\tau,\textbf{R})=\sum_{\nu} \tilde{A}(\textbf{R}) e^{\rmi 2\pi \nu T\tau}$ does not depend on $\tau$. In fact, $\tilde{A}(\textbf{R})$ acts like a \emph{classical} bosonic field with free energy 
\begin{align}
 \nonumber F[\tilde{A}] ={}& \int \mbox{d}^2R_\perp \int_0^{L_z}\mbox{d}Z\Biggl[ \nabla \tilde{A}_{\mu i}^*\cdot \nabla \tilde{A}_{\mu i} \\
\nonumber &+ (\gamma-1)(\partial_i \tilde{A}_{\mu i}^*)(\partial_j \tilde{A}_{\mu j}) \\
\label{bos2} &+ \alpha\ \mbox{tr}(\tilde{A}^\dagger \tilde{A})+\sum_{a=1}^5 \bar{\beta}_{a} \tilde{I}_a\Biggr].
\end{align}
Here we normalized the order parameter field $\tilde{A}$ such that the coefficient of the leading gradient term is unity.

The symmetry group of the free energy in Eq. (\ref{bos2}) is $\tilde{G}$. 
The difference in the number of spatial indices $i$ of $A_{\mu i}(\textbf{R})$, which is $2$, and the number of spatial coordinates $\textbf{R}=(X,Y,Z)$, which is $3$, makes this bosonic theory rather non-trivial. In the following, we call the situation with $0<L_z<\infty$ the \emph{quasi-2D limit}. For $L_z=0$, which is equivalent to ignoring the $Z$-dependence of $A_{\mu i}(\textbf{R})$, we recover the $\text{SO}(3)_{\rm S} \times \text{SO}(2)_{\rm L} \times \text{U}(1)$ theory studied by Jones, Love, Moore \cite{Jones:1976zz} when $\gamma=1$. We refer to the situation without $Z$-coordinate as the \emph{2D limit}.

\begin{table}[t!]
\begin{tabular}{|c|c|c|c|c|c|}
\hline
 \ length \ & $k_{\rm F}^{-1}$ &$\lambda_{T_{\rm c}}$ & $\xi_{T=0}$ & $L_z$ & $L_x$ \\
\hline
 & $0.1\ \text{nm}$ &  $30\ \text{nm}$  & $70\ \text{nm}$ & $1000\ \text{nm}$ & $1\ \text{mm}$ \\
\hline\hline
 \ temperature \  & $T_{\rm F}$ & $T_{\rm c}$ & $E_\xi/k_{\rm B}$ & $E_z/k_{\rm B}$ & $E_x/k_{\rm B}$ \\
\hline
 & $8\ \text{K}$ & $1\ \text{mK}$ & $0.01\ \text{mK}$  & $0.1\ \mu \text{K}$ & $10^{-7}\ \mu \text{K}$\\
\hline
\end{tabular}
\label{TabUnits}
\caption{\textbf{Typical experimental scales.} We collect typical experimental length and temperature scales for settings as in Ref. \cite{PhysRevLett.124.015301}. Here $k_{\rm F}$ is the Fermi momentum and $T_{\rm F}=\frac{\hbar^2k_{\rm F}^2}{2mk_{\rm B}}$ the Fermi temperature with $m$ the mass of a He-3 atom. The thermal wave length  is $\lambda_T=\sqrt{\frac{2\pi\hbar^2}{mk_{\rm B}T}}$, evaluated here at the phase transition with $T=T_{\rm c}$. The coherence length at zero temperature is $\xi_{T=0}$ and $E_\xi=\frac{\hbar^2}{2m\xi^2}$ the associated energy scale. The spatial extensions in the (confined) z-direction and (unconfined) x-direction are $L_z$ and $L_x$, respectively, corresponding to the energy scales $E_z = \frac{\hbar^2}{2mL_z^2}$ and $E_x = \frac{\hbar^2}{2mL_x^2}$.}
\end{table}

The spontaneous breaking of the continuous symmetry group $\tilde{G}$ leads to the presence of several Goldstone modes in the superfluid phase. They correspond to gapless fluctuations, i.e. fluctuations that do not cost energy in the limit of infinite wavelengths, which are prone to destroy the superfluid long-range order. These gapless modes lead to divergences in diagrammatic contributions to the free energy that require a regularization and renormalization scheme.

In the following, we apply the weak-coupling momentum-shell RG \cite{PhysRevB.97.064504} to incorporate the effect of fluctuations of the order parameter, especially the associated Goldstone modes. For this purpose, we restrict the momentum integration in Feynman diagrams due to order parameter fluctuations to a momentum shell via
\begin{align}
 \label{bos4} \int_{\textbf{q}_\perp}^\prime (\dots) = \frac{1}{2\pi} \int_{\Omega/b}^\Omega\mbox{d}q_\perp\ q_\perp(\dots).
\end{align}
Here, $\textbf{q}_\perp=(q_x,q_y)$ is the two-dimensional momentum in the unconfined xy-plane, $\Omega\sim \sqrt{T}$ is the ultraviolet cutoff of the bosonic theory, and $b>1$ is the RG flow parameter. Since all momenta inside the momentum shell have $q_\perp>0$, the infrared singularity at $q_\perp  =0$ is avoided. Successively increasing $b\to \infty$ amounts to including all order parameter fluctuations.

Within the RG picture, the coefficients of the free energy, in particular the quartic couplings $\beta_1(b),\dots,\beta_5(b)$, depend on the RG flow parameter $b$. Any fixed valued of $b>1$ corresponds to a system with typical momentum scale $k_{\rm typ} \sim \Omega/b$ and typical wavelength of excitations  $\lambda_{\rm typ} \sim (\Omega/b)^{-1}$. As $b\to \infty$, $k_{\rm typ}\sim 0$ and $\lambda_{\rm typ}\sim \infty$, and fluctuations on all length scales are included. In any realistic system, there is a maximal length scale set by the system size---in our case this is $L_x\sim L_y$. The RG needs to be stopped at $b \sim L_x\Omega$, since fluctuations cannot have wavelengths that are larger than the system size. Characteristic length scales of nanoscale-confined experiments with He-3 are listed in Tab. \ref{TabUnits}.

\subsection{2D-limit}

We first discuss the perturbative  RG in the 2D-limit, where the $Z$-dependence of $\tilde{A}_{\mu i}(X,Y,Z)$ is neglected. This assumes that fluctuations in the z-direction are not important at the phase transition. However, the 2D-limit captures the contributions from the divergent, gapless modes. For this reason, it shares the same qualitative features as the more elaborate RG flow in the quasi-2D regime, which is discussed below. The one-loop flow equations for the rescaled couplings
\begin{align}
 \label{bos4b} \beta_a(b) = \frac{\bar{\beta}_a(b)}{8\pi \Omega^2}
\end{align}
have the form
\begin{align}
 \label{bos4c} \frac{\mbox{d}\beta_a}{\mbox{d}\ln b} = 2\beta_a - \frac{\mathcal{C}_a(\gamma)}{\gamma^2}, 
\end{align}
where $\mathcal{C}_a(\gamma)$ is a quadratic form of the couplings. We explicitly have
\begin{widetext}
\begin{align}
 \nonumber \frac{\mbox{d}\beta_1}{\mbox{d}\ln b} = 2\beta_1 -\frac{1}{\gamma^2}\Bigl[{}& 24(\kappa^2+2\kappa+2)\beta_1^2+12(\kappa+2)^2\beta_1\beta_2+4(5\kappa^2+8\kappa+8)\beta_1\beta_3+8(\kappa^2+2\kappa+2)\beta_1\beta_4\\
 \label{bos5} &+24(\kappa^2+2\kappa+2)\beta_1\beta_5+4\kappa^2\beta_2\beta_5+2(5\kappa^2+16\kappa+16)\beta_3\beta_5+2\kappa^2\beta_4\beta_5+3\kappa^2\beta_5^2\Bigr]\\
 \nonumber \frac{\mbox{d}\beta_2}{\mbox{d}\ln b} = 2\beta_2 -\frac{1}{\gamma^2}\Bigl[{}&4(\kappa^2+8\kappa+8)\beta_1^2+16(\kappa^2+2\kappa+2)\beta_1\beta_2+4\kappa^2\beta_1\beta_3+4\kappa^2\beta_1\beta_4+4\kappa^2\beta_1\beta_5\\
 \nonumber &+2(17\kappa^2+40\kappa+40)\beta_2^2+4(9\kappa^2+16\kappa+16)\beta_2\beta_3+4(11\kappa^2+20\kappa+20)\beta_2\beta_4\\
 \nonumber &+24(\kappa^2+2\kappa+2)\beta_2\beta_5+2(3\kappa^2+4\kappa+4)\beta_3^2+16(\kappa^2+\kappa+1)\beta_3\beta_4+2(5\kappa^2+8\kappa+8)\beta_3\beta_5\\
 \label{bos6} &+12(\kappa^2+2\kappa+2)\beta_4^2+2(5\kappa^2+8\kappa+8)\beta_4\beta_5+(\kappa^2+8\kappa+8)\beta_5^2\Bigr]\\
 \nonumber\frac{\mbox{d}\beta_3}{\mbox{d}\ln b} = 2\beta_3 
 -\frac{1}{\gamma^2}\Bigl[{}&4\kappa^2\beta_1^2+4\kappa^2\beta_1\beta_3+4(\kappa^2+8\kappa+8)\beta_1\beta_4+4(\kappa^2+8\kappa+8)\beta_1\beta_5+2\kappa^2\beta_2^2\\
 \nonumber &+12(\kappa+2)^2\beta_2\beta_3+4\kappa^2\beta_2\beta_4+2(7\kappa^2+8\kappa+8)\beta_3^2+16(\kappa^2+5\kappa+5)\beta_3\beta_4\\
  \label{bos7} &+2(\kappa^2+8\kappa+8)\beta_3\beta_5+4\kappa^2\beta_4^2+2\kappa^2\beta_4\beta_5+\kappa^2\beta_5^2\Bigr]\\
 \nonumber \frac{\mbox{d}\beta_4}{\mbox{d}\ln b} = 2\beta_4  -\frac{1}{\gamma^2}\Bigl[{}&4\kappa^2\beta_1^2+4(\kappa^2+8\kappa+8)\beta_1\beta_3+4\kappa^2\beta_1\beta_4+4(5\kappa^2+8\kappa+8)\beta_1\beta_5+2\kappa^2\beta_2^2+4\kappa^2\beta_2\beta_3\\
 \nonumber &+12(\kappa+2)^2\beta_2\beta_4+2(3\kappa^2+20\kappa+20)\beta_3^2+16(\kappa^2+\kappa+1)\beta_3\beta_4+2\kappa^2\beta_3\beta_5\\
 \label{bos8} &+4(3\kappa^2+10\kappa+10)\beta_4^2+2(\kappa^2+8\kappa+8)\beta_4\beta_5+(17\kappa^2+32\kappa+32)\beta_5^2\Bigr]\\
 \nonumber \frac{\mbox{d}\beta_5}{\mbox{d}\ln b} = 2\beta_5 -\frac{1}{\gamma^2}\Bigl[{}&8\kappa^2\beta_1\beta_2+8(\kappa+2)^2\beta_1\beta_3+16(\kappa^2+2\kappa+2)\beta_1\beta_4+16(\kappa^2+3\kappa+3)\beta_2\beta_5\\
 \label{bos9} &+4(3\kappa^2+4\kappa+4)\beta_3\beta_5+4(7\kappa^2+16\kappa+16)\beta_4\beta_5+6(\kappa+2)^2\beta_5^2\Bigr].
\end{align}
\end{widetext}
Here we denote $\kappa=\gamma-1$. These equations for $\gamma\neq 1$ constitute a central result of this work. For $\gamma=1$, they agree with the flow equations derived in Ref. \cite{Jones:1976zz}. 
The initial conditions for the RG flow at $b=1$ are given by the mean-field values
\begin{align}
 \label{bos10}(\beta_1(1),\dots,\beta_5(1)) = \beta_{2,\rm MF}\Bigl(-\frac{1}{2},1,1,1,-1\Bigr).
\end{align}

The flow equation for $\gamma$ in the 2D limit has been computed in Ref. \cite{Jones:1976zz} and reads
\begin{align}
 \label{bos11} \frac{\mbox{d}\gamma}{\mbox{d}\ln b}  =\frac{16}{3} (1-\gamma)\Bigl(\frac{1}{\gamma^2}+3\Bigr)f_\gamma,
\end{align}
where $f_\gamma>0$ is a positive definite function given by
\begin{align}
 \nonumber f_\gamma={}& 12 \beta_1^2+2\beta_1\beta_2+8\beta_1\beta_3+2\beta_1\beta_4+6\beta_1\beta_5\\
 \nonumber &+\frac{13}{2}\beta_2^2+4\beta_2\beta_3+7\beta_2\beta_4+5\beta_2\beta_5+8\beta_3^2\\
 \label{bos12} &+4\beta_3\beta_4+\frac{13}{2}\beta_4^2+5\beta_4\beta_5+\frac{15}{2}\beta_5^2.
\end{align}
The associated flow of $\gamma(b)$ has an infrared stable fixed point at $\gamma_\star =1$ where the kinetic term in the action simplifies considerably. However, the initial value from mean-field theory is
\begin{align}
 \label{bos13} \gamma(1) = \gamma_{\rm MF} = 3.
\end{align}
Consequently, in the initial stages of the flow with $b\gtrsim 1$, the parameter $\gamma >1$ needs to be taken into account. However, the flow of $\gamma(b)$ has only a quantitative effect on the running of the quartic couplings $\beta_a(b)$.

\begin{figure}
    \centering
   \includegraphics[width=8cm]{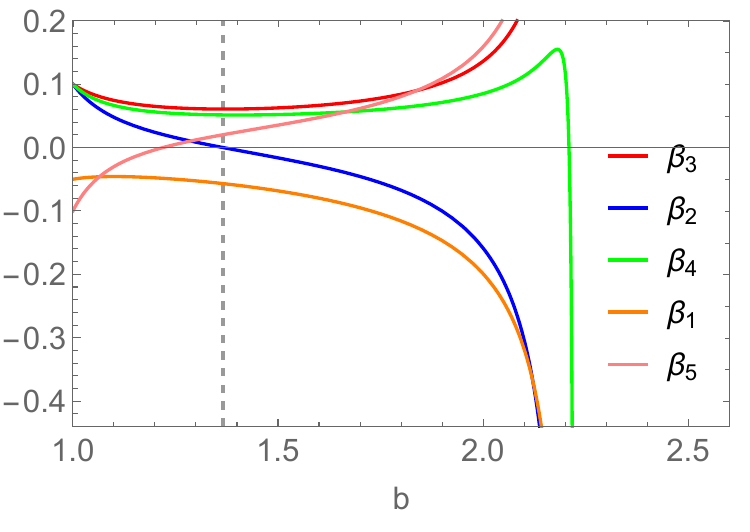}
   \includegraphics[width=4.2cm]{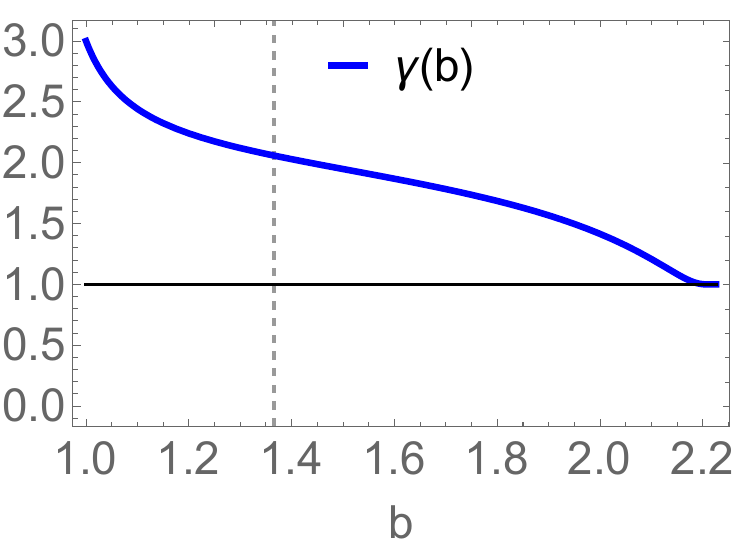}
   \includegraphics[width=4.2cm]{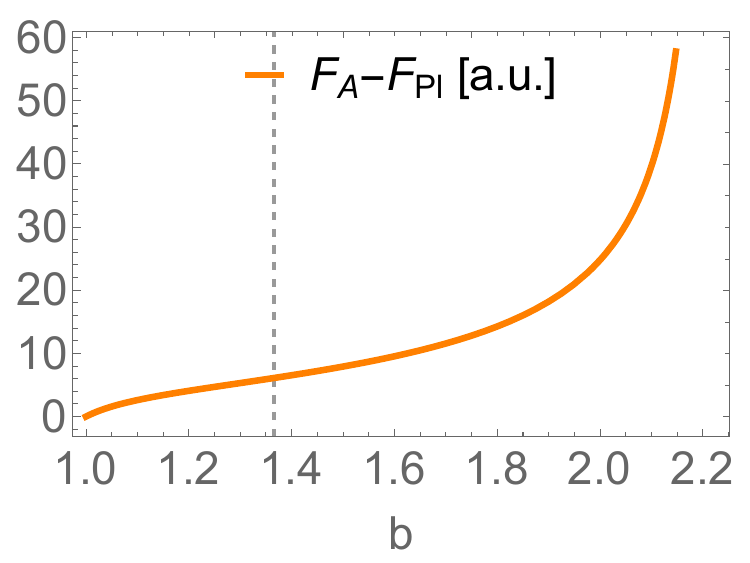}
    \caption{\textbf{RG flow in the 2D limit.} \emph{Top.} Fluctuations of the order parameter lead to a scale-dependence or ``running" of the quartic coefficients $\beta_a(b)$, where $b$ represents the typical wavelength of fluctuations according to $\lambda \sim b \lambda_T$ (with $\lambda_T\propto\sqrt{T}$ the thermal wavelength). The RG is such that the flow enters an unphysical regime with $\beta_2(b)<0$ at some value of $b\sim b_0$, indicated by the dashed vertical line, which corresponds to a breakdown of the perturbative RG. \emph{Bottom left.} The kinetic coefficient $\gamma(b)$ quickly flows from its mean-field value $\gamma(0)=3$ to the infrared stable fixed point at $\gamma_\star =1$. The dashed vertical lines in both bottom panels again indicates $b_0$. \emph{Bottom right.} The free energy of the A-phase, $F_A$, and the planar phase, $F_{\rm pl}$, are such that the planar phase is energetically favored in the initial stage of the flow, when the perturbative RG is applicable. We plot their difference versus $b$ in arbitrary units.}
    \label{Fig2D}
\end{figure}

Solving the RG flow equations in the 2D-limit, we find that the five quartic couplings $\beta_a(b)$ for $b>1$ quickly deviate from the fixed ratios of the mean-field initial conditions in Eq. (\ref{bos10}), see Fig. \ref{Fig2D}. This lifts the accidental energetic degeneracy of the A-phase and planar phase. The coefficients $\beta_a(b)$ in the early stages $b  \sim 1$ are such that the planar phase is energetically favored over the A-phase.

We observe that the flow quickly enters an unphysical regime at $b\sim b_{0}$, where either one of the couplings $\beta_a(b_0)$ diverges (violating the assumptions of weak coupling), or the free energy becomes unbounded from below due to $\beta_2(b_0)=0$ (invalidating the expansion of the free energy to quartic order in the field). We discuss below how this behavior should be interpreted, but the perturbative RG equation can no longer be applied in this regime. We find that for all $b \leq b_{0}$, the planar phase is energetically favored over the A-phase. We also observe that these features of the RG flow are independent of whether we incorporate the full flow equation for $\gamma(b)$, or simply set $\gamma(b)$ to the constant values of either 3 or 1.

\subsection{Quasi-2D limit}

In the quasi-2D limit, fluctuations of the order parameter in the confined z-direction are taken into account. The RG flow depends on $L_z$ through the additional parameter \cite{PhysRevA.93.063631}
\begin{align}
 \label{bos14} \tilde{L}_z = k_{\rm typ}L_z = \frac{\Omega L_z}{b}.
\end{align}
The physical meaning of the dimensionless number $\tilde{L}_z$ is the effective length of the z-dimension seen by a fluctuation with typical wavevector $k_{\rm typ}=\Omega/b$. For instance, the experimental value of $L_z \sim 500\text{nm}$ is large for a fluctuation with wavelength $\lambda \sim 10\text{nm}$, but would appear small for fluctuations with $\lambda \sim 10 \mu\text{m}$. With $k_{\rm typ} = 2\pi/\lambda$, these exemplary  parameter sets correspond to $\tilde{L}_z=600$ and $\tilde{L}_z=0.6$, respectively. In experiment, a typical value of $L_z\Omega$ is
\begin{align}
 \label{bos15} (L_z\Omega)_{\rm exp} \sim 30,
\end{align}
hence $\tilde{L}_z\gg 1$ in the early stages of the flow with $b\sim 1$.

Our mathematical treatment of the order parameter fluctuations of $\tilde{A}_{\mu i}(X,Y,Z)$ in the quasi-2D limit is as follows. For the Fourier transform of $\tilde{A}_{\mu i}$ with wave vector $\textbf{q}=(q_x,q_y,q_z)$, we assume that the momentum components $\textbf{q}_\perp = (q_x,q_y)$ are part of a continuum of modes. The relative spacing between these discrete momenta would be of order $ \pi/L_{x}\sim \pi/L_y$, which we can safely neglect. On the other hand, we assume that fluctuations in the z-direction are quantized according to
\begin{align}
 \label{bos16} q_z = \frac{\pi n}{L_z},
\end{align}
with integer $n=1,2,\dots$, implementing Dirichlet boundary conditions. The lowest possible excitation has wavevector $\textbf{q}=(0,0,\pi/L_z)$, corresponding to $n=1$. This is the gapless Goldstone mode in the confined geometry. The associated nonzero kinetic energy can be compensated by an appropriate choice of the quadratic coefficient $\alpha= - \pi^2/L_z^2$. On the other hand, fluctuations or excitations with $n>1$ are energetically suppressed.

The different quantization schemes applied to $\textbf{q}_\perp$ and $q_z$ introduce a spatial anisotropy that is reflected in the RG flow equations. For $\gamma\neq 1$, there is an additional anisotropy that stems from the fluctuation propagator, because $\gamma-1$ only multiplies the momentum components $q_x$ and $q_y$. We assume the flow of the coefficient $\gamma$ to be determined by the 2D-limit from Eq. (\ref{bos11}). For $\gamma=1$, the flow of the quartic couplings has the simple form
\begin{align}
 \label{bos17} \frac{\mbox{d}\beta_a}{\mbox{d}\ln b} = 2\beta_a -  \mathcal{L}(\tilde{L}_z)\ \mathcal{C}_a(1),
\end{align}
with $\mathcal{C}_a(1)=\mathcal{C}_a(\gamma=1)$ from Eq. (\ref{bos4c}), and the scaling function
\begin{align}
 \label{bos18} \mathcal{L}(\tilde{L}_z) = \frac{\tilde{L}_z^3}{4}\frac{-2+\eta\ \coth(\eta) + \eta^2/\sinh^2(\eta)}{\eta^4}>0
\end{align}
with $\eta=(\tilde{L}_z^2-\pi^2)^{1/2}$. The 2D-limit from Eq. (\ref{bos4c}) with $\gamma=1$ is obtained by setting $\mathcal{L}(\tilde{L}_z)=1$. Similarly, the flow equations in the quasi-2D regime with $\gamma\neq 1$ can be computed analytically, see App. \ref{AppFlowQuart}.

\begin{figure}
    \centering
   \includegraphics[width=8cm]{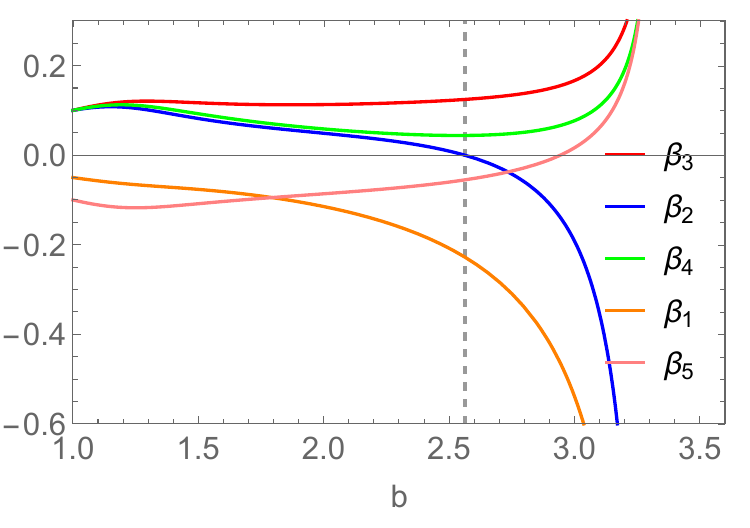}
   \includegraphics[width=4.2cm]{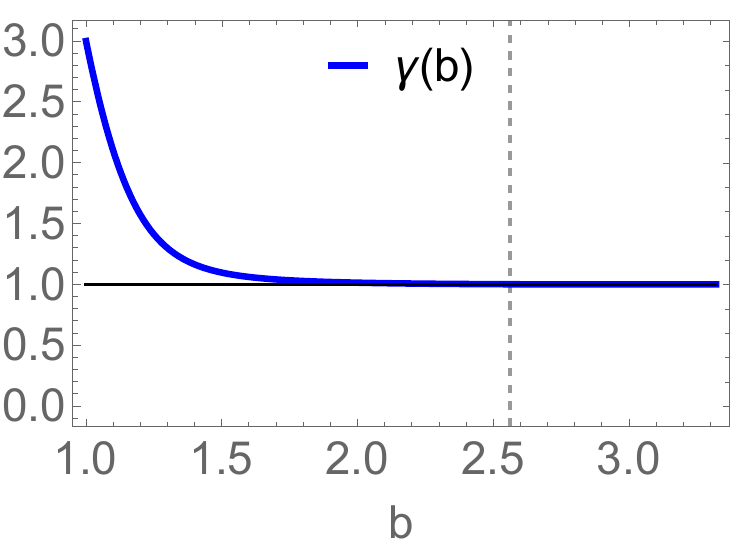}
   \includegraphics[width=4.2cm]{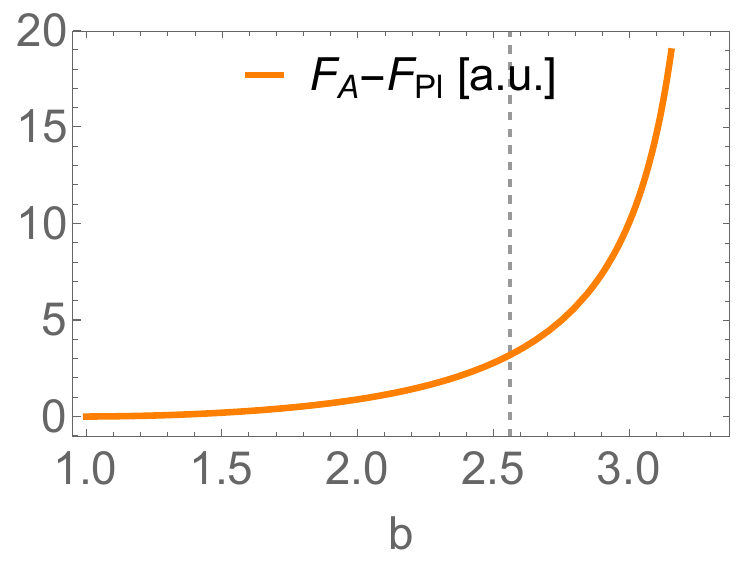}
    \caption{\textbf{RG flow in the quasi-2D limit.} The RG flow in the quasi-2D regime, plotted here for the experimental value $L_z\Omega=30$, differs only quantitatively from the 2D limit shown in  Fig. \ref{Fig2D} with the same choice of labels.  In particular, the planar phase is again energetically favored over the A-phase. On the other hand, the value of $b_0$ is slightly larger in the quasi-2D regime.}
    \label{FigQ2D}
\end{figure}

In Fig. \ref{FigQ2D} we show the solution to the RG flow equations in the experimental regime $L_z\Omega \sim 30$. The flow of the coefficient $\gamma$ is incorporated via Eq. (\ref{bos11}), although a quantitatively similar result is obtained by setting $\gamma(b)=1$ in the flow equations for the quartic couplings, since $\gamma(b)$ is attracted to the value $\gamma_\star=1$. We observe that the running of the quartic couplings $\beta_a(b)$ in the quasi-2D limit is qualitatively similar to the result in the 2D-limit. The RG flow quickly enters an unphysical regime with negative $\beta_2(b)$ at $b_0 \sim 1$, although $b_0$ is slightly larger in the quasi-2D regime than the 2D-limit. For all $b<b_0$, the planar phase is energetically favored over the A-phase.

\subsection{Interpretation of the perturbative RG flow}

In the study of both classical and quantum phase transitions, second-order phase transitions are commonly associated with stable infrared fixed points of the RG flow. Such fixed points capture the experimentally observed scale invariance and allow us to determine experimentally observed critical exponents. This procedure has been applied and benchmarked successfully for superfluid order parameters in three-dimensional bosonic XY-models, or models containing fermions in the Gross--Neveu--Yukawa class. More involved scenarios such as the Berezinskii--Kosterlitz--Thouless transition of the two-dimensional XY-model still fall into this scheme when allowing for lines of fixed points in parameter space.

On the other hand, the absence of a stable infrared fixed point, or more generally a runaway flow of couplings (if the initial conditions are not within the basin of attraction of a stable infrared fixed point), is commonly interpreted as sign of a fluctuation-induced first-order phase transition. For instance, the fluctuations of the photon gauge field in an ordinary superconductor have this effect. Depending on the size of the parameter $b_0>1$, which characterizes the breakdown of the RG flow, the fluctuation-induced first-order transition is considered to be either weak or strong. For this note that we associate the value of $b_0$ with a typical momentum of fluctuations $k_0 \sim \Omega/b_0$, or typical energy $E_0 \sim \Omega^2/b_0 = T/b_0^2$. For a first-order phase transition at $T=T_{\rm c}$, the typical energy is the induced jump of the order parameter $E_0\sim \Delta_{\rm 1st}$, hence
\begin{align}
 \label{int2} \Delta_{\rm 1st} \sim \frac{T_{\rm c}}{b_0^2}.
\end{align}
If $b_0\gg1$, then the induced jump is small and the first-order transition is considered to be weak. In contrast, if $b_0 \gtrsim 1$, then the first-order transition is considered to be strong and should be experimentally detectable.

In the present case of nanoscale confined He-3, with runaway RG flow and  $b_0\gtrsim 1$, we would be lead to the conclusion that the transition from the normal to the superfluid phase is a strong fluctuation-induced first-order transition. Similarly, for bulk He-3 in 3D, the RG flow computed by Jones, Love, Moore does not yield a stable infrared fixed point and so would also predict that the transition in 3D helium-3 is not of second order. This conclusion, however, is in strong contrast to the experimental finding of a seemingly smooth second-order transition without jump in the superfluid density \cite{PhysRevLett.74.4667,PhysRevLett.124.015301}. One may explain this by a weak first-order transition with $b_0\gg 1$, assuming  that the experiments do not resolve energy differences or temperature scales as low as $E_0 \sim T_{\rm c}/b_0^2$. But given that $b_0 \sim 1$ in the study of the RG flow of the He-3 order parameter in various geo\-metries, theory and experiment clearly disagree on the nature of the superfluid phase transition.

To explain this discrepancy, one may, of course, argue that perturbative RG does not apply to He-3, because the system is strongly coupled even at low pressures. However, another, perhaps additional explanation for the short-coming strikes us to be the following. The fixed points of perturbative RG are computed close to the non-interacting, Gaussian fixed point through expansion in some small parameter $\vare\ll 1$, which here is the difference from 4 dimensions. An infrared-stable fixed point is characterized by purely negative eigenvalues of the stability matrix at the fixed point; furthermore, the dimension of the stability matrix is equal to the number of quartic couplings. For the Wilson--Fisher fixed point of classical $\text{O}(N)$-models, there is only one coupling constant, and its stability eigenvalue at the interacting fixed point is, indeed, proportional to $-\vare$, whereas it is $+\vare$ at the Gaussian fixed point.  In the present matrix model, on the other hand, there are \emph{five} quartic couplings and the eigenvalues of the Gaussian fixed point are $(\vare,\vare,\vare,\vare,\vare)$. It seems rather unlikely that a single one-loop calculation can change the sign of five eigenvalues to yield an infrared-stable interacting fixed point that could describe the second-order phase transition. As such, it may also hardly be surprising that many other tensor field theories do not find stable fixed points in the perturbative RG, see e.g. Refs. \cite{Jones:1976zz,PhysRevB.23.3549,PhysRevD.29.338,PhysRevB.38.4916,1995PhLA..208..161A,2001NuPhB.607..605P,PhysRevB.63.140414,PhysRevA.93.051603,PhysRevB.97.064504}

\section{Summary and Outlook}
In conclusion, we have studied the superfluid transition of helium-3 under uniaxial nanoscale confinement. The key observation made in this work is the reduction of the $3\times 3$, $p$-wave, triplet matrix order parameter in 3D to a $3\times 2$ matrix under confinement.
This has important physical consequences, such as guaranteeing a nodal quasiparticle spectrum, regardless of the exact form of the ground state obtained from minimizing the Ginzburg--Landau free energy functional.

To illustrate the emergence of the $3\times 2$ matrix order parameter, we first analyzed the Cooper problem of two interacting atoms in the presence of a Fermi surface. Under confinement, in order to satisfy the Dirichlet boundary conditions, the Cooper pair must acquire a non-zero center-of-mass momentum. This momentum acts as an obstacle to forming Cooper pairs near the Fermi surface, reducing the region in momentum space where Cooper pairs can form. This anisotropic reduction of momentum space raises the energy of a Cooper pair in a $p_z$ orbital more than in a $p_x$ or $p_y$ orbital. Thus the $p_x$ and $p_y$ orbitals are energetically favored and more likely to condense at the phase transition.

While the Cooper problem provides tremendous physical intuition, to capture many-body effects and obtain the precise thermodynamic ground state, we derive the Landau free energy under confinement within the mean-field approximation. We observed the $3\times 3$ matrix order parameter splitting into two, a $3\times 2$ matrix $\tilde{A}$ and a $3\times 1$ matrix $A_z$, with $\tilde{A}$ possessing a higher critical temperature. Hence, in the vicinity of the Fermi liquid to superfluid phase transition, only $\tilde{A}$ remains relevant. Remarkably, the mean-field Landau free energy for the reduced order parameter can be minimized exactly. The ground state manifold is categorized into two classes of states. The first consists of time-reversal broken states that are symmetry related to the A phase, and the second class of states is related by symmetry to the time-reversal symmetric planar phase. While they are degenerate on a mean-field level, the A-phase has a lower free energy once strong coupling corrections are included.

To complement the mean-field theory and strong-coupling analysis, we also performed a perturbative RG analysis to investigate the effects of order parameter fluctuations at the putative second-order phase transition. The interplay between the anisotropic geometry and the matrix nature of the order parameter leads to a rather intricate bosonic field theory. Nevertheless, we have been able to analytically determine the RG flow equations both in the presence of confinement (for $0\leq L_z<\infty$) and non-standard kinetic terms (for $\gamma \neq 1$). Integration of the RG flow yields a runaway flow of the quartic couplings, which indicates a fluctuation-induced first-order transition, with the planar phase energetically preferred over the A-phase. We argued that the perturbative RG might fall short to capture the superfluid transition in helium-3---even in the 3D case---and that the straightforward interpretation of the outcome is hence questionable.

Given the five quartic couplings $\beta_1,\dots,\beta_5$ of the Landau free energy $F^{(4)}[\tilde{A}]$ in any approximation, it is a very nontrivial problem to determine the ensuing minimum of $F^{(4)}[\tilde{A}]$. This is because the order parameter $\tilde{A}$ has many components and the symmetry group $\tilde{G}$ relating equivalent states is also large. Nonetheless, we were able to exactly minimize the mean-field free energy by showing that it is the sum of three positive terms and subsequently searching for configurations that minimize each term. We found that only two symmetry-inequivalent ground states exist, the A-phase and the planar phase, which break or preserve time-reversal symmetry, respectively. Beyond the mean-field approximation, the minimization of $F^{(4)}[\tilde{A}]$ is more complicated. According to Michel's theorem, one can look for ground states within all stationary states that are invariant under subgroups of $\tilde{G}$. Here we only considered the A-phase and planar phase, since these are the energetically degenerate ground states dictated by mean-field theory, and leave the more detailed search for future work.

Our analysis supplies a simple and analytically tractable picture for superfluid orders of He-3 under confinement. The crucial observation is that despite the microscopic interaction remaining unchanged under confinement, the boundary conditions necessitate a non-zero center-of-mass momentum, reducing the symmetry of the Hamiltonian. Thus the order parameter is modified and needs to be reclassified accordingly. In fact, a similar analysis can be performed for helium-3 in a cylinder confined in the radial direction. In this geometry, the order parameter again splits into $\tilde{A}$ and $A_z$, but $A_z$ has a higher critical temperature. This is consistent with previous work simulating the phase diagram with maximal pair breaking boundary conditions \cite{PhysRevB.92.144515}.

\section*{Ackowledgements}
The authors thank Albion Arifi, Anffany Chen, John Davis, Igor Herbut, Lukas Janssen, Joseph Maciejko, Subrata Mandal, Alex Shook, and Pramodh Senarath Yapa for inspiring discussions. CS acknowledges support through the Natural Sciences and Engineering Research Council of Canada (NSERC) Discovery Grant RGPAS-2020-00064 and the Pacific Institute for the Mathematical Sciences CRG PDF Fellowship Award. AA and IB acknowledge support through the University of Alberta startup fund UOFAB Startup Boettcher.  IB acknowledges funding from the NSERC Discovery Grants RGPIN-2021-02534 and DGECR2021-00043.

\begin{appendix}

\section{Derivation of the Ginzburg-Landau free energy}\label{app: MF coefficients}

In this appendix, we provide more details regarding the derivation of the Ginzburg--Landau free energy Eq. \eqref{eq: G-L} in the main text. 

Our starting point is the action Eq. \eqref{eq: MF exact action} from the main text,
 \begin{equation}
F[A]=\frac{\beta \mathcal{V}}{g}\sum_{\mu ,j}\bar{A}_{\mu j}A_{\mu j}-\frac{1}{2}\sum_{k}\tr\ln \mathcal{G}^{-1}(k).
\end{equation}
For notational compactness, we omit the $\vq$ dependence in $A_{\mu j}$ and $\mathcal{G}^{-1}$. As the order parameter is small in the vicinity of the phase transition, we expand the action in powers of $A_{\mu j}$. To this end, we write the inverse Green function as
\begin{equation}
    \mathcal{G}^{-1}(k)=\mathcal{G}^{-1}_{0}(k)+\hat{\Delta}(\vk),
\end{equation}
where
\begin{align}
    \mathcal{G}_{0}^{-1}(\vq,\vk)&=\begin{bmatrix}
		ik_n-\xi_{\vk_+}&0\\
		0&ik_n+\xi_{\vk_-}
	\end{bmatrix},\\
 \hat{\Delta}(\vk)&=\begin{bmatrix}
			0&-\Delta(\vk)\\
			-\Delta^*(\vk)&0
	\end{bmatrix}.
\end{align}
Here $\mathcal{G}_{0}^{-1}(k)$ is the inverse Green function in the normal state. The trace-log term in the action can be expanded in powers of $\Delta(\vk)$ as
\begin{align}
&\tr \ln \mathcal{G}^{-1}(k)=\tr \ln \left[\mathcal{G}_{0}^{-1}(,k)(1+\mathcal{G}_{0}(k)\hat{\Delta}(\vk)) \right]\nonumber\\
&=\tr \ln \mathcal{G}_{0}^{-1}(k)+\tr \ln \left[1+\mathcal{G}_{0}(k)\hat{\Delta}(\vk)\right]\nonumber\\
	&=\tr\ln \mathcal{G}_{0}^{-1}(k)+ \sum_{m=1}^\infty \frac{(-1)^{m+1}}{m}\tr \left[(\mathcal{G}_0(k)\hat{\Delta}(\vk))^m\right]\nonumber\\
	&=\tr\ln \mathcal{G}_{0}^{-1}(k)- \sum_{m=1}^\infty \frac{1}{2m}\tr \left[(\mathcal{G}_{0}(k)\hat{\Delta}(\vk))^{2m}\right]\nonumber\\
	&=\tr\ln \mathcal{G}^{-1}_{0}(k)-\sum_{m=1}^\infty \frac{1}{m}\frac{\tr\left[(\Delta^*(\vk)\Delta(\vk))^m\right]}{(ik_n-\xi_{\vk_+})^m(ik_n+\xi_{\vk_-})^m}.
\end{align}
The first term is a constant that does not depend on $\Delta(\vk)$; as such it is inessential for our purposes and will be ignored. The free energy can thus be expressed as the perturbative expansion
\begin{align}\label{eq: free energy expansion}
	&F[A]=\frac{\beta \mathcal{V}}{g}\sum_{\mu j}A^*_{\mu j}A_{\mu j}\nonumber\\
 &+ \frac{1}{2}\sum_{k} \sum_{m=1}^\infty \frac{1}{m}\frac{\tr \left[(\Delta^*(\vk)\Delta(\vk))^{m}\right]}{(ik_n-\xi_{\vk_+})^m(ik_n+\xi_{\vk_-})^m}.
\end{align}

\subsection{Quadratic coefficients}
The quadratic term in Eq. \eqref{eq: free energy expansion} is
\begin{align}\label{eq: S2 A}
	F^{(2)}[A]= \beta\mathcal{V}\sum_{\mu,j_1j_2}A^*_{\mu j_1}\left[\frac{\delta_{j_1j_2}}{ g}-K_{j_1 j_2} \right]A_{\mu j_2},
\end{align}
where we abbreviate
\begin{equation}
	K_{j_1j_2}\equiv -\frac{1}{\beta\mathcal{V}}\sum_k \frac{1}{(ik_n-\xi_{\vk_+})(ik_n+\xi_{\vk_-})}\frac{3k_{j_1}k_{j_2}}{k_{\rm F}^2}.
\end{equation}
Evaluating the Matsubara sum according to
\begin{align}
	K_{j_1j_2}=\frac{1}{\mathcal{V}}\sum_{\vk}\frac{1-n_{\rm F}(\xi_{\vk_+})-n_{\rm F}(\xi_{\vk_-})}{\xi_{\vk_+}+\xi_{\vk_-}}\frac{3k_{j_1}k_{j_2}}{k_{\rm F}^2},
\end{align}
where $n_{\rm F}(x)=(e^{\beta x}+1)^{-1}$ is the Fermi-Dirac distribution. As $|\vq|\ll  |\vk|\sim k_{\rm F}$, we expand $K$ in powers in $\vq$. To second order we have
\begin{align}
	&\frac{1-n_{\rm F}(\xi_{\vk_+})-n_{\rm F}(\xi_{\vk_-})}{\xi_{\vk_+}+\xi_{\vk_-}}=\nonumber\\
 &\frac{1-2n_{\rm F}(\xi_\vk)}{2\xi_\vk}- \frac{n_{\rm F}^{\prime\prime}(\xi_\vk)\epsilon_\vk\epsilon_\vq\cos^2\theta_{\vk-\vq}}{2\xi_\vk} +\mathcal{O}(q^3),
\end{align}
where $\theta_{\vk-\vq}$ is the angle between $\vk$ and $\vq$. This expansion is a good approximation when $\beta\mu q/k_{\rm F}\ll 1$, which is generally the case. In this expression we have omitted terms that are odd in $\xi_\vk$ as they would be zero once the radial integral is performed. We now take the continuum limit. As the momentum sum is over a thin shell about the Fermi surface, we can replace the sum $\frac{1}{\mathcal{V}}\sum_\vk\mapsto \int \frac{\d^3k}{(2\pi)^3}\mapsto N(0)\int_{-\Lambda}^\Lambda \d \xi\int \frac{\d\Omega}{4\pi} $. Making this substitution, the kernel becomes
\begin{align}\label{eq: K kernel before angular integration}
	&K_{j_1j_2}=\delta_{j_1j_2}N(0)\underbrace{\int_0^{\Lambda}\d \xi \frac{1}{\xi}\tanh\left(\frac{\beta\xi}{2}\right)}_{\to\ln \left(\frac{2e^\gamma}{\pi}\frac{\Lambda}{k_{\rm B} T}\right)}\\
 &- N(0)\mu\epsilon_\vq\underbrace{\int_{-\Lambda}^{\Lambda} \d \xi \frac{n_{\rm F} ^{\prime\prime}(\xi)}{2\xi}}_{\to\beta^2\frac{7\zeta(3)}{4\pi^2}}\int\frac{\d\Omega}{4\pi}\cos^2\theta_{\vk-\vq}\frac{3k_{j_1}k_{j_2}}{k_{\rm F}^2}.
\end{align}
Here the radial integrals were evaluated in the limit $\beta\Lambda\to \infty$.
The angular integral is
\begin{equation}
    \int\frac{\d\Omega}{4\pi}\cos^2\theta_{\vk-\vq}\frac{3k_{j_1}k_{j_2}}{k_{\rm F}^2}=\frac{\delta_{j_1j_2}}{5}\times\begin{cases}
		1&j_1=x,y\\
		3&j_1=z
	\end{cases}
\end{equation}
Hence,
\begin{align}\label{eq: K}
    K_{j_1j_2}&=\delta_{j_1j_2}N(0)\ln \left(\frac{2e^\gamma}{\pi}\frac{\Lambda}{k_{\rm B} T} \right)\nonumber\\
    &-\frac{7\zeta(3)}{20\pi^2}\frac{N(0)\mu\epsilon_\vq}{(k_{\rm B}T)^2}\times\begin{cases}
        1&j_1=x,y\\
        3&j_1=z
    \end{cases}.
\end{align}
Combining Eqs. \eqref{eq: S2 A} and \eqref{eq: K}, we obtain Eqs. \eqref{eq: S2}, \eqref{eq: alphaxy}, and \eqref{eq: alphaz} from the main text.

\subsection{Quartic coefficients}
The quartic coefficients can be computed analogously. The quartic contribution to the free energy in Eq. \eqref{eq: free energy expansion} is
\begin{equation}
	F^{(4)}[A]=\frac{1}{4}\sum_k \frac{\tr\left[\Delta^*(\vk)\Delta(\vk)\Delta^*(\vk)\Delta(\vk) \right]}{(ik_n-\xi_{\vk_+})^2(ik_n+\xi_{\vk_-})^2}.
\end{equation}
Let us write
\begin{equation}
    F^{(4)}[A]\equiv \beta \mathcal{V}\sum_{\mu,j} A^*_{\mu_1 j_1}A^*_{\mu_2 j_2}L_{\mu,j}A_{\mu_3 j_3}A_{\mu_4 j_4},
\end{equation}
where the kernel
\begin{align}\label{eq: L kernel}
    L_{\mu j}=\frac{1}{4\beta \mathcal{V}}\sum_k \frac{\tr(\sigma_{\mu_1}\sigma_{\mu_2}\sigma_{\mu_3}\sigma_{\mu_4})}{(ik_n-\xi_{\vk_+})^2(ik_n+\xi_{\vk_-})^2}\frac{9k_{j_1}k_{j_2}k_{j_3}k_{j_4}}{k_{\rm F}^4}.
\end{align}
Here we have used the short-hand notation $\mu$ to denote the four indices $\mu_1,\mu_2,\mu_3$, and $\mu_4$, and similarly for $j$. As $\tilde{A}$ has a higher critical temperature than $A_{\rm z}$, we focus solely on the $3\times 2 $ matrix by allowing the indices $j_1$, $j_2$, $j_3$, and $j_4$ to only run over $x$ and $y$. 

The evaluation of the kernel Eq. \eqref{eq: L kernel} can be split into three parts. First, the trace over the four spin matrices evaluates to 
\begin{align}
    \frac{1}{2}\tr(\sigma_{\mu_1}\sigma_{\mu_2}\sigma_{\mu_3}\sigma_{\mu_4})&=\delta_{\mu_1\mu_3}\delta_{\mu_2\mu_4}+\delta_{\mu_1\mu_4}\delta_{\mu_2\mu_3}\nonumber\\
    &-\delta_{\mu_1\mu_2}\delta_{\mu_3\mu_4}.
\end{align}
Second, evaluating the Matsubara sum and expanding the resulting expression to second order in $\vq$ we obtain
\begin{align}
\nonumber	 &\frac{1}{\beta}\sum_{ik_n}\frac{1}{(ik_n-\xi_{\vk_+})^2(ik_n+\xi_{\vk_-})^2}\\
\nonumber  &=\frac{2[1-n_F(\xi_{\vk_+})-n_F(\xi_{\vk_-})]}{(\xi_{\vk_+}+\xi_{\vk_-})^3}+\frac{n_F^\prime(\xi_{\vk_+})+n_F^\prime(\xi_{\vk_-})}{(\xi_{\vk_+}+\xi_{\vk_-})^2}
\\
\nonumber &=
	\frac{[1-2n_F(\xi_\vk)]+2\xi_\vk n_F^\prime(\xi_\vk)}{4\xi_\vk^3}\\
 &-\frac{\epsilon_\vk \epsilon_\vq \cos^2\theta_{\vk,\vq}}{4\xi_\vk^3}[n_F^{\prime\prime}(\xi_\vk)-\xi_\vk n_F^{\prime\prime\prime}(\xi_\vk)]+\mathcal{O}(q^3).
\end{align}
As before, we have ignored terms that are odd in $\xi_\vk$. Finally, the angular integrals are
\begin{align*}
	\int\frac{\d\Omega}{4\pi}\frac{9k_{j_1}k_{j_2}k_{j_3}k_{j_4}}{k_{\rm F}^4}&=\frac{3}{5}( \delta_{j_1j_2}\delta_{j_3j_4}+\delta_{j_1j_3}\delta_{j_2j_4}\\
 &\hspace{0.8cm}+\delta_{j_1j_4}\delta_{j_2j_3})\\
 \int\frac{\d\Omega}{4\pi}\frac{9 k_z^2 k_{j_1}k_{j_2}k_{j_3}k_{j_4}}{k_F^6}&=\frac{3}{35}(\delta_{j_1j_2}\delta_{j_3j_4}+\delta_{j_1j_3}\delta_{j_2j_4}\\
 &\hspace{0.8cm}+\delta_{j_1j_4}\delta_{j_2j_3} ).
\end{align*}
Combining everything, we obtain
\begin{align}
    &L_{\mu j}=\beta_2\left(-\frac{1}{2}\tilde{I}_1+\tilde{I}_2+\tilde{I}_3+\tilde{I}_4-\tilde{I}_5\right),
\end{align}
where
\begin{align}
    \beta_2= \frac{3N(0)}{20}\bigg[&\int_{-\Lambda}^\Lambda \d \xi \frac{[1-2n_{\rm F}(\xi)+2 \xi n_{\rm F}^\prime (\xi)]}{\xi^3} \\
    &-\frac{\mu \epsilon_\vq}{7}\int_{-\Lambda}^\Lambda\d\xi \frac{n_{\rm F}^{\prime\prime}(\xi)-\xi n_{\rm F}^{\prime\prime\prime}(\xi)}{\xi^3}\bigg].
\end{align}
Here we have used that $\epsilon_\vk\simeq \mu$ for $\mu\gg \Lambda$. In the limit $\beta\Lambda\to \infty$, the integrals are
\begin{align}
    \int_{-\Lambda}^\Lambda \d \xi \frac{[1-2n_{\rm F}(\xi)+2 \xi n_{\rm F}^\prime (\xi)]}{\xi^3} &\to\frac{7\zeta(3)}{2\pi^2}\beta^2\\
    \int_{-\Lambda}^\Lambda\d\xi \frac{n_{\rm F}^{\prime\prime}(\xi)-\xi n_{\rm F}^{\prime\prime\prime}(\xi)}{\xi^3}&\to \frac{93\zeta(5)}{4\pi^4}\beta^4.
\end{align}
Therefore,
\begin{equation}
    \beta_2=\frac{21\zeta(3)}{40\pi^2}\frac{N(0)}{(k_{\rm B}T)^2}-\frac{279\zeta(5)}{560\pi^4}\frac{N(0)\mu}{(k_{\rm B}T)^4}\epsilon_\vq,
\end{equation}
which is Eq. \eqref{eq: beta2} of the main text.

\section{Exact minimization of the mean-field free energy}\label{app: energy minimization}

In order to minimize $F^{(4)}$, it is convenient to express the orbital part of the order parameter $\tilde{A}$ in the angular momentum basis. We define the $3\times 2$ complex matrix
\begin{align}\label{eq: A to B basis transformation}
\tilde{B}^\prime &\equiv \tilde{A}^\prime U_{\rm L},\ U_{\rm L}=\frac{1}{\sqrt{2}}\begin{pmatrix}
        1&1\\
        -i&i
    \end{pmatrix}.
\end{align}
The order parameter $\tilde{B}$ can be written in matrix form as
\begin{equation}
    \tilde{B}^\prime=\begin{pmatrix}
        \vec{S}&\vec{T}
    \end{pmatrix},
\end{equation}
[where $\vec{S}$ and $\vec{T}$ are three-component complex column vectors. Under a spin rotation, the vectors $\vec{S}$ and $\vec{T}$ transform in the vector representation of $\SO(3)_{\rm S}$, but under an orbital rotation from $\text{SO}(2)_{\rm L}$ by an angle $\phi$, we have
\begin{align}
\vec{S}&\mapsto e^{i\phi}\vec{S},\\ \vec{T}&\mapsto e^{-i\phi}\vec{T}.
\end{align}
In other words, $\vec{S}$ ($\vec{T}$) transforms in the $m=+1$ $(-1)$ angular momentum representation of $\SO(2)_{\rm L}$. Time-reversal acts as $\vec{S}\mapsto \vec{T}^*$ and $\vec{T}\mapsto \vec{S}^*$. The basis transformation Eq. \eqref{eq: A to B basis transformation} preserves the normalization condition Eq. \eqref{eq: A normalization}, namely
    \begin{equation}\label{eq: B normalization}
        \tr \tilde{B}^{\prime \dagger} \tilde{B}^\prime=|\vec{S}|^2+|\vec{T}|^2=1.
    \end{equation}In this notation, the quartic invariants read
\begin{align}
    \tilde{I}_1&=4|\vec{S}\cdot \vec{T}|^2\\
    \tilde{I}_2&= (|\vec{S}|^2+|\vec{T}|^2)^2=1\\
    \tilde{I}_3&= 2(|\vec{S}|^2|\vec{T}|^2+|\vec{S}\cdot \vec{T}^*|^2)\\
    \tilde{I}_4&=|\vec{S}|^4+|\vec{T}|^4+2 |\vec{S}\cdot \vec{T}^*|^2\\
    \tilde{I}_5&= |\vec{S}^2|^2+|\vec{T}^2|^2+2|\vec{S}\cdot \vec{T}|^2.
\end{align}

We show these inequalities for normalized $\tilde{B}$, implying they are valid for the non-normalized as well (yielding an overall prefactor $|\Delta_0|^4$). To prove Eq. \eqref{eq: inequality 31}, we first write, using the Binet–Cauchy identity,
\begin{align}
    \tilde{I}_3-\frac{\tilde{I}_1}{2}
    &=2\left(|\vec{S}|^2|\vec{T}|^2+(\vec{S}\times \vec{S}^*)\cdot (\vec{T}^*\times \vec{T})\right).
\end{align}
Sparating $\vec{S}$ and $\vec{T}$ into real and imaginary parts, $\vec{S}=\vec{S}_1+i\vec{S}_2$ and $\vec{T}=\vec{T}_1+i\vec{T}_2$, where $\vec{S}_1,\vec{S}_2,\vec{T}_1,\vec{T}_2\in \mathbb{R}^3$, we have
\begin{align}
    \tilde{I}_3-\frac{\tilde{I}_1}{2}&=2(S_1^2+S_2^2)(T_1^2+T_2^2)\nonumber\\
    &+8S_1S_2T_1T_2 (\hat{\vec{S}}_1\times \hat{\vec{S}}_2)\cdot (\hat{\vec{T}}_1\times \hat{\vec{T}}_2).
\end{align}
Here $S_1=|\vec{S}_1|$ denotes the magnitude of $\vec{S}_1$ and $\hat{\textbf{S}}_1=\textbf{S}_1/S_1$, and likewise for $S_2$,
$T_1$, and $T_2$. Because of the normalization condition, Eq. \eqref{eq: B normalization}, it is helpful to introduce angular variables $\theta,\phi_S,\phi_T\in [0,\pi/2]$ to parameterize the magnitudes: $S_1=\cos\theta \cos\phi_S$, $S_2=\cos\theta \sin\phi_S$, $T_1=\sin\theta\cos\phi_T$, and $T_2=\sin\theta\sin\phi_T$. In this parameterization,
\begin{align}
\tilde{I}_3-\frac{\tilde{I}_1}{2}&= \frac{\sin^22\theta}{2}\bigg[1+\sin2\phi_S \sin 2 \phi_T \nonumber\\
&\hspace{2.1cm}\cdot (\hat{\vec{S}}_1\times \hat{\vec{S}}_2)\cdot (\hat{\vec{T}}_1\times \hat{\vec{T}}_2)\bigg] \label{eq: I3-I1}\\
&\geq \frac{\sin^22\theta}{2}(1-\sin2\phi_S \sin 2 \phi_T )\\
&\geq 0.
\end{align}
The second inequality, Eq. \eqref{eq: inequality 45}, can be proven in much the same way; the result is
\begin{equation}\label{eq: I4-I5}
    \tilde{I}_4-\tilde{I}_5=4(\vec{S}_1\times \vec{S}_2+ \vec{T}_1\times \vec{T}_2)^2\geq 0.
\end{equation}

The inequality \eqref{eq: inequality 45}, in fact, remains true in the 3D case when the order parameter is a $3\times 3$ matrix $A$, namely
\begin{align}
 I_4 - I_5\geq 0
\end{align}
for
$I_4=\tr[(AA^\dagger)^2]$ and $
I_5=\tr[(AA^\dagger)(AA^\dagger)^*]$. However, inequality \eqref{eq: inequality 31} is generally not satisfied in 3D. Take, as a counterexample, the order parameter of the B phase $A=\diag (1,1,1)/\sqrt{3}$, for which it can be easily verified that $I_3-I_1/2=-1/6$.

The fact that both inequalities \eqref{eq: inequality 31} and \eqref{eq: inequality 45} can be satisfied by the $3\times 2$ order parameter makes the minimization of the mean-field free energy analytically tractable. For this we classify all superfluid ground states by seeking order parameters that saturate both bounds. From Eq. \eqref{eq: I3-I1}, there are two instances when $\tilde{I}_3-\tilde{I}_1/2=0$. The first is when
\begin{align}\label{eq: 31 constraint 1}
    \theta&=0 \text{  or }\frac{\pi}{2},
\end{align}
and the second when
\begin{align}\label{eq: 31 constraint 2}
\phi_S&=\phi_T=\frac{\pi}{4},&
(\hat{\vec{S}}_1\times \hat{\vec{S}}_2)\cdot (\hat{\vec{T}}_1\times \hat{\vec{T}}_2)&=-1.
\end{align}
By \eqref{eq: I4-I5}, $\tilde{I}_4-\tilde{I}_5=0$ when
\begin{equation}\label{eq: 45 constraint}
    \vec{S}_1\times \vec{S}_2=-\vec{T}_1\times \vec{T}_2.
\end{equation}
Based on these conditions, the ground state manifold can be categorized into two classes:
\begin{enumerate}
    \item States that satisfy Eqs. \eqref{eq: 31 constraint 1} and \eqref{eq: 45 constraint}. In this class, one of $\vec{S}$ or $\vec{T}$ is zero and the non-zero vector, say $\vec{S}$, satisfies $\vec{S}_1\parallel \vec{S}_2$. In other words, $\vec{S}_1$ and $\vec{S}_2$ must point in the same direction but their relative magnitude is unconstrained. 

    The states within this class are related to one another by $\tilde{G}$-transformations. Without the loss of generality, suppose $\vec{S}\neq 0 $ and $\vec{T}=0$. By an $\SO(3)_S$ transformation, $\vec{S}_1$ can be brought to the form $(0,0,\cos\phi_S)^T$.  The ground state and normalization constraints then demand $\vec{S}_2=\pm (0,0,\sin\phi_S)^T$. Performing the $\text{U}(1)$-transformation $\vec{S}\to e^{\mp i\phi_S}\vec{S}$, $\vec{T}\to e^{\mp i\phi_S}\vec{T}$, we obtain
    \begin{align}
        \vec{S}_1&=\begin{pmatrix}
            0\\
            0\\
            1
        \end{pmatrix},\ \vec{S}_2=\vec{T}_1=\vec{T}_2=0.
    \end{align}
Any state in this class can be brought to this canonical form through $\tilde{G}$-transformations.  In the Cartesian basis, this representative state takes the form
       \begin{align}
        \tilde{A}^\prime=\frac{1}{\sqrt{2}}\begin{pmatrix}
            0&0\\
            0&0\\
            1&i
        \end{pmatrix},
    \end{align}
    which is the A phase. The case with $\vec{S}=0$ and $\vec{T}\neq 0$ is related by time-reversal symmetry. Note that states in this class necessarily break time-reversal symmetry.

    \item States that satisfy Eqs \eqref{eq: 31 constraint 2} and \eqref{eq: 45 constraint}. The magnitudes are constrained to $S_1=S_2=T_1=T_2=1/2$ and the directions satisfy the three conditions $\hat{\vec{S}}_1\perp \hvec{S}_2$, $\hvec{T}_1\perp \hvec{T}_2$, and $\hat{\vec{S}}_1\times \hvec{S}_2$ antiparallel to $\hvec{T}_1\times \hvec{T}_2$. 
    
    States in this class are also all related by $\tilde{G}$-transformations. By an $\SO(3)_{\rm S}$ rotation, $\hvec{S}_1$ can be chosen to be $\hvec{S}_1=(1,0,0)^T$. The vector $\hvec{S}_2$ must be perpendicular to $\hvec{S}_1$, and, by a spin rotation about $\hvec{S}_1$, we can choose $\hvec{S}_2=(0,-1,0)^T$. As for the vectors $\hvec{T}_1$ and $\hvec{T}_2$, in order for $\hvec{T}_1\perp \hvec{T}_2$ and $\hvec{S}_1\times \hvec{S}_2$ to be antiparallel to $\hvec{T}_1\times \hvec{T}_2$, we must have $\hvec{T}_1=\cos\beta \hvec{S}_1+\sin\beta \hvec{S}_2$ and $\hvec{T}_2=\sin\beta \hvec{S}_1-\cos\beta \hvec{S}_2$ for some $\beta\in[0,2\pi)$. The parameter $\beta$, however, can be eliminated by performing the $\text{U}(1)$-transformation $\vec{S}\mapsto e^{-i\beta/2}\vec{S}$, $\vec{T}\to e^{-i\beta/2}\vec{T}$ and $\SO(2)_{\rm L}$ transformation $\vec{S}\mapsto e^{i\beta/2}\vec{S}$, $\vec{T}\mapsto e^{-i\beta/2}\vec{T}$. After this sequence of transformations, we obtain
    \begin{align}
      \hvec{S}_1&=\begin{pmatrix}
          1\\
          0\\
          0
      \end{pmatrix},\
      \hvec{S}_2=\begin{pmatrix}
          0\\
          -1\\
          0
      \end{pmatrix},\\
      \hvec{T}_1&=\hvec{S}_1,\ \hvec{T}_2=-\hvec{S}_2.
    \end{align}
   In other words, all the states in this class are  $\tilde{G}$-related to this time-reversal symmetric state. Reverting to the Cartesian basis, this representative state reads
   \begin{equation}
       \tilde{A}^\prime=\frac{1}{\sqrt{2}}\begin{pmatrix}
           1&0\\
           0&1\\
           0&0
       \end{pmatrix},
   \end{equation}
   which is the planar phase.
\end{enumerate}

To summarize, our weak coupling calculation predicts two classes of ground states that are not related by symmetry. The states in the first class are related to the A phase by $\tilde{G}$-transformations, and to the planar phase in the second.

\section{Perturbative RG equations}\label{AppBRG}

Our derivation of the perturbative RG equations is based on using the (bosonic) one-loop effective action as a generating functional for the  one-loop beta functions of the quartic couplings. This is an efficient way of deriving the one-loop flow of models with matrix order parameters and many quartic couplings, where otherwise a substantial number of Feynman diagrams would have to be considered. The technique is explained in detail in Appendix D of Ref. \cite{PhysRevB.97.064504}.

\subsection{One-loop effective action}

We parametrize the effective free energy functional for the order parameter at a given RG scale $b$ through the ansatz
\begin{align}
 \nonumber F[\tilde{A}(\textbf{R})] ={}& \int_{\textbf{R}}\Bigl[ (\partial_k \tilde{A}_{\mu i}^*)(\partial_k \tilde{A}_{\mu i})\\
 \label{pert1} &+ (\gamma-1) (\partial_i \tilde{A}^*_{\mu i})(\partial_j \tilde{A}_{\mu j}) + U(\tilde{A}(\textbf{R}))\Bigr].
\end{align}
Note that
\begin{align}
 \mbox{tr}[(\partial_k\tilde{A}^\dagger)(\partial_k \tilde{A})] = (\partial_k \tilde{A}_{\mu i}^*)(\partial_k \tilde{A}_{\mu i}).
\end{align}
The effective potential $U(\tilde{A})$ contains all terms in the free energy density that are allowed by symmetry and independent of derivatives of $\tilde{A}(\textbf{R})$. We limit the perturbative study to the RG-relevant couplings close to the phase transition and make the ansatz
\begin{align}
 \label{pert2} U(\tilde{A}) = -\alpha \ \mbox{tr}(\tilde{A}^\dagger \tilde{A})+\sum_{a=1}^5 \bar{\beta}_a \tilde{I}_a.
\end{align}
In Eqs. (\ref{pert1}) and (\ref{pert2}), the couplings $\gamma, \alpha, \bar{\beta}_a$ depend on the RG scale parameter $b$. 

The one-loop correction to the effective potential is given by the "trace-log" formula
\begin{align}
 \label{pert3} \delta U(\tilde{A}) = \frac{1}{2}\int_{\textbf{q}}^\prime \ln G^{-1}(\textbf{q},\tilde{A}),
\end{align}
where $G^{-1}$ is the Fourier transform of the inverse propagator in the presence of an \emph{arbitrary} background field $\tilde{A}(\textbf{R})$, denoted $\mathcal{G}^{-1}$, evaluated at a \emph{constant} background field $\tilde{A}(\textbf{R})=\tilde{A}$. It is defined through
\begin{align}
  \label{pert4} G^{-1}(\textbf{q},\tilde{A}) = \begin{pmatrix} \mathcal{G}^{-1}_{\tilde{A}\tilde{A}}(\textbf{q}) & \mathcal{G}^{-1}_{\tilde{A}\tilde{A}^*}(\textbf{q}) \\ \mathcal{G}^{-1}_{\tilde{A}^*\tilde{A}}(\textbf{q}) & \mathcal{G}^{-1}_{\tilde{A}^*\tilde{A}^*}(\textbf{q})\end{pmatrix}\Bigr|_{\tilde{A}(\textbf{R})=\tilde{A}},
\end{align}
where each block is an $3\times 2$-matrix defined by the functional derivatives
\begin{align}
 \label{pert5} \frac{\delta^2 F}{\delta \tilde{A}_{\mu i}(\textbf{R})\delta \tilde{A}_{\nu j}(\textbf{R}')} = [\mathcal{G}^{-1}_{\tilde{A}\tilde{A}}(\textbf{R},\tilde{A}(\textbf{R}))]_{\mu i,\nu j}\delta^{(3)}(\textbf{R}-\textbf{R}'),\\
 \label{pert6} \frac{\delta^2 F}{\delta \tilde{A}_{\mu i}(\textbf{R})\delta \tilde{A}_{\nu j}^*(\textbf{R}')} = [\mathcal{G}^{-1}_{\tilde{A}\tilde{A}^*}(\textbf{R},\tilde{A}(\textbf{R}))]_{\mu i,\nu j}\delta^{(3)}(\textbf{R}-\textbf{R}'),
\end{align}
with analogous formulas for $\mathcal{G}^{-1}_{\tilde{A}^*\tilde{A}}$ and $\mathcal{G}^{-1}_{\tilde{A}^*\tilde{A}^*}$. After setting $\tilde{A}(\textbf{R})\to \tilde{A}$ to a constant, the Fourier transform $\textbf{R}\to \textbf{q}$ can be computed. (This would yield a complicated convolution for a non-constant order parameter field.) It is important though to keep the value of $\tilde{A}$ arbitrary in $G^{-1}(\textbf{q},\tilde{A})$ so that derivatives of $\delta U(\tilde{A})$ with respect to $\tilde{A}$ can be computed. After these derivatives have been obtained, we insert a vanishing background field $\tilde{A} = 0$. For this we define
\begin{align}
\label{pert7} G_0^{-1}(\textbf{q}) &= G^{-1}(\textbf{q},\tilde{A}=0)
\end{align}
and the associated inverse matrix $G_0(\textbf{q})$. The choice of momentum shell integration $\int_{\textbf{q}}^\prime$ in Eq. (\ref{pert3}) is discussed below in Eq. (\ref{pert20}).

To determine the one-loop corrections to the five couplings $\bar{\beta}_a$ we define
\begin{align}
 \label{EqJ1} J_1 &= \frac{\partial^4 \delta U}{\partial \tilde{A}_{11}\partial \tilde{A}_{11}^*\partial \tilde{A}_{11}\partial \tilde{A}_{11}^*}\Biggr|_{\tilde{A}=0},\\
 J_2 &= \frac{\partial^4 \delta U}{\partial \tilde{A}_{11}\partial \tilde{A}_{11}^*\partial \tilde{A}_{22}\partial \tilde{A}_{22}^*}\Biggr|_{\tilde{A}=0},\\
J_3 &= \frac{\partial^4 \delta U}{\partial \tilde{A}_{11}\partial \tilde{A}_{11}^*\partial \tilde{A}_{12}\partial \tilde{A}_{12}^*}\Biggr|_{\tilde{A}=0},\\
 J_4 &= \frac{\partial^4 \delta U}{\partial \tilde{A}_{11}\partial \tilde{A}_{11}^*\partial\tilde{A}_{21}\partial \tilde{A}_{21}^*}\Biggr|_{\tilde{A}=0},\\
 \label{pert8} J_5 &= \frac{\partial^4 \delta U}{\partial \tilde{A}_{12}\partial \tilde{A}_{11}^*\partial \tilde{A}_{32}\partial \tilde{A}_{31}^*}\Biggr|_{\tilde{A}=0},
\end{align}
and find
\begin{align}
 J_1 &=4(\delta \bar{\beta}_1+\delta \bar{\beta}_2+\delta \bar{\beta}_3+\delta\bar{\beta}_4+\delta \bar{\beta}_5),\\
 J_2 &= 2\delta \bar{\beta}_2,\\
 J_3 &= 2(\delta \bar{\beta}_2+\delta \bar{\beta}_4+\delta \bar{\beta}_5),\\
 J_4 &= 2(\delta \bar{\beta}_2+\delta \bar{\beta}_3+\delta \bar{\beta}_4),\\
 \label{pert9} J_5 &=2\delta \bar{\beta}_3.
\end{align}
Here we denote the one-loop correction to $\bar{\beta}_a$ by $\delta \bar{\beta}_a$. This is inverted by
\begin{align}
 \delta \bar{\beta}_1 &= \frac{1}{4}(J_1-2J_3-2J_5),\\
 \delta \bar{\beta}_2 &= \frac{1}{2}J_2,\\
 \delta \bar{\beta}_3 &= \frac{1}{2}J_5,\\
 \delta \bar{\beta}_4 &= \frac{1}{2}(-J_2+J_4-J_5),\\
 \label{pert10} \delta \bar{\beta}_5 &= \frac{1}{2}(J_3-J_4+J_5).
\end{align}
These expressions are quadratic in the couplings $\{\bar{\beta}_a\}$ and the RG flow equations for the $\{\beta_a\}$ can be derived from them, as we explain in the next section.

To compute the derivatives of $\delta U$ in Eqs. (\ref{EqJ1})-(\ref{pert8}) in practice we make repeated use of
\begin{align}
 \label{pert12} \frac{\mbox{d}}{\mbox{d}t} M(t) = - M(t) \Bigl(  \frac{\mbox{d}}{\mbox{d}t} M(t)^{-1}\Bigr) M(t)
\end{align}
for any matrix $M(t)$ that depends on a parameters $t$. This yields
\begin{align}
 \nonumber &\frac{\partial^4 \delta U}{\partial \tilde{A}_{\mu i}^*\partial \tilde{A}_{\nu j}^*\partial \tilde{A}_{\sigma k}\partial \tilde{A}_{\tau l}}\Bigr|_{\tilde{A}=0} \\
 \nonumber &= -\frac{1}{2}\mbox{tr}\int_{\textbf{q}}^\prime G_0 \frac{\partial^2 \mathcal{V}}{\partial \tilde{A}_{\mu i}^*\partial \tilde{A}_{\nu j}^*} G_0 \frac{\partial^2 \mathcal{V}}{\partial \tilde{A}_{\sigma k}\partial \tilde{A}_{\tau l}}\\
 \nonumber & -\frac{1}{2}\mbox{tr}\int_{\textbf{q}}^\prime G_0 \frac{\partial^2 \mathcal{V}}{\partial \tilde{A}_{\mu i}^*\partial \tilde{A}_{\sigma k}} G_0 \frac{\partial^2 \mathcal{V}}{\partial \tilde{A}_{\nu j}^*\partial \tilde{A}_{\tau l}} \\
\label{pert13} &-\frac{1}{2}\mbox{tr}\int_{\textbf{q}}^\prime G_0 \frac{\partial^2 \mathcal{V}}{\partial \tilde{A}_{\mu i}^*\partial \tilde{A}_{\tau l}}G_0 \frac{\partial^2 \mathcal{V}}{\partial \tilde{A}_{\nu j}^*\partial \tilde{A}_{\sigma k}}.
\end{align}
Here $\mathcal{V}(\tilde{A})$ is derived from the ansatz for $F$ in Eqs. (\ref{pert1}) and (\ref{pert2}). We have 
\begin{align}
 \label{pert14}   G^{-1}(\textbf{q},\tilde{A}) = G_0^{-1}(\textbf{q}) + \mathcal{V}(\tilde{A})
\end{align}
with
\begin{align}
 \label{pert14b} \mathcal{V}(\tilde{A})_{\mu i,\nu j} &= \begin{pmatrix} \frac{\partial^2U}{\partial \tilde{A}_{\mu i}\partial \tilde{A}_{\nu j}} & \frac{\partial^2U}{\partial \tilde{A}_{\mu i}\partial \tilde{A}_{\nu j}^*} \\ \frac{\partial^2U}{\partial \tilde{A}_{\mu i}^*\partial \tilde{A}_{\nu j}} & \frac{\partial^2U}{\partial \tilde{A}_{\mu i}^*\partial \tilde{A}_{\nu j}^*} \end{pmatrix},\\
 \nonumber  G_0^{-1}(\textbf{q})_{\mu i,\nu j} &=  \delta_{\mu\nu} \Biggl[ \Bigl(q^2\delta_{ij}-q_iq_j\Bigr)+\gamma q_iq_j\Biggr]\begin{pmatrix} 0 & 1 \\ 1 & 0 \end{pmatrix}.\\
 \label{pert15}
\end{align}
These are $2(3\times 2) \times 2(3\times 2) = 18 \times 18$ matrices. The vertex term $\mathcal{V}(\tilde{A})$ in Eq. (\ref{pert14}) contributes terms proportional to the quartic couplings $\{\beta_a\}$ when computing the derivatives in Eqs. (\ref{pert8}). It depends on the particular ansatz used for $U(\tilde{A})$ in Eq. (\ref{pert2}).

In the momentum integrals, we parametrize $\textbf{q}$ as
\begin{align}
 \label{pert16} \textbf{q} = \begin{pmatrix} \textbf{q}_\perp \\ q_z \end{pmatrix} =  \begin{pmatrix} q_x \\ q_y \\ q_z \end{pmatrix} = \begin{pmatrix} q \cos \phi \\ q \sin\phi \\ q_z \end{pmatrix},
\end{align}
where $\textbf{q}_\perp=(q_x,q_y)$ is the momentum in the (unconfined) two-dimensional xy-plane, and $q_z$ is the momentum component in the (confined) z-direction. Under confinement, $q_z$ is quantized to values 
\begin{align}
 \label{pert17} q_z = \frac{\pi}{L_z} n.
\end{align}
The integration over $q_z$ is replaced by a Matsubara-type sum
\begin{align}
 \label{pert18} \frac{1}{L_z} \sum_{n\geq 1}.
\end{align}
We assume $q_z\ll p_\perp$ so that typically only the $n=1$ term contributes significantly. The integral over the two-momentum $\textbf{q}_\perp$ is given by
 \begin{align}
 \label{pert19}\frac{1}{(2\pi)^2} \int_{\Omega/b}^\Omega \mbox{d}q_\perp\ q_\perp \int_0^{2\pi} \mbox{d}\phi,
\end{align}
where $q_\perp=|\textbf{q}_\perp|$ is limited to the momentum shell $\Omega/b \leq q_\perp \leq \Omega$ with ultraviolet cutoff $\Omega\sim\sqrt{T}$. We combine these consideration into the prescription
\begin{align}
\nonumber &\int_{\textbf{q}}^\prime f(\textbf{q}_\perp,q_z) \\
\label{pert20}  &:=  \frac{1}{L_z(2\pi)^2} \sum_{n\geq 1}\int_{\Omega/b}^\Omega \mbox{d}q_\perp\ q_\perp \int_0^{2\pi} \mbox{d}\phi\ f(\textbf{q}_\perp,\pi n/L_z)
\end{align}
for an arbitrary function $f(\textbf{q}_\perp,q_z)$.

\subsection{From one-loop corrections to RG flow equations}

In this section we explain how the one-loop corrections $\delta \bar{\beta}_a$ to the quartic couplings $\bar{\beta}_a$ translate into a set of coupled RG flow equations. To simplify the presentation, we consider a model that has only one quartic coupling $\bar{\lambda}$ with a scaling dimension of 2. (This means that if $k$ is a momentum scale, then $\bar{\lambda}/k^2$ is dimensionless in appropriate natural units.) We explain at the end of the section how this generalizes to models with several couplings.

The one-loop correction is assumed to have the form
\begin{align}
 \delta \bar{\lambda} = \mathcal{C}  \bar{\lambda}^2 \int_{\textbf{q}_\perp}^\prime \frac{1}{q_\perp^4},
\end{align}
where $\mathcal{C}$ is a dimensionless number. In fact, the $\bar{\lambda}$ that appears on the right-hand side is the coupling $\bar{\lambda}(b)$ at the scale $b=1$, so we write
\begin{align}
 \delta \bar{\lambda} = \mathcal{C}  \bar{\lambda}(1)^2 \int_{\textbf{q}_\perp}^\prime \frac{1}{q_\perp^4}.
\end{align}
The momentum-shell integral is evaluated in two dimensions and given by
\begin{align}
\int_{\textbf{q}_\perp}^\prime \frac{1}{q_\perp^4} = \frac{b^2-1}{4\pi \Omega^2}.
\end{align}
We now define the \emph{renormalized coupling} at scale $b$ via
\begin{align}
 \bar{\lambda}(b) &= b^2\Bigl( \bar{\lambda}(1) + \delta\bar{\lambda}\Bigr)\\
 &=b^2\Bigl( \bar{\lambda}(1) +\mathcal{C}\bar{\lambda}(1)^2 \frac{b^2-1}{4\pi \Omega^2}\Bigr).
\end{align}
The scale derivative with respect to $\ln b$ yields
\begin{align}
 \frac{\mbox{d}\bar{\lambda}(b)}{\mbox{d}\ln b} &= b \frac{\mbox{d}\bar{\lambda}(b)}{\mbox{d}b}\\
 &=2\bar{\lambda}(b) +\frac{\mathcal{C}}{2\pi \Omega^2} [b^2\bar{\lambda}(1)]^2.
\end{align}
In the loop-correction term, we now self-consistently replace
\begin{align}
 b^2\bar{\lambda}(1) \to \bar{\lambda}(b).
\end{align}
This is justified, for instance, if the coupling corrections are small, or if $b$ is close to unity. We arrive at
\begin{align}
 \frac{\mbox{d}\bar{\lambda}(b)}{\mbox{d}\ln b}
 &=2\bar{\lambda}(b) +  \frac{\mathcal{C}}{2\pi \Omega^2} \bar{\lambda}(b)^2.
\end{align}
This is the one-loop RG flow equation for the coupling $\bar{\lambda}$. Since in many situations the overall prefactor of $\bar{\lambda}$ does not play a role, it is convenient to introduce the rescaled coupling
\begin{align}
 \lambda(b) = \frac{\bar{\lambda}(b)}{2\pi \Omega^2}
\end{align}
to arrive at the equation
\begin{align}
 \frac{\mbox{d}\lambda}{\mbox{d}\ln b}
 &=2\lambda +  \mathcal{C} \lambda^2.
\end{align}
Note that $\lambda$, in contrast to $\bar{\lambda}$, is a dimensionless number. The case of several coupling constants $\bar{\lambda}_i$ is obtained by replacing $\mathcal{C}\bar{\lambda}^2$ with the quadratic form
\begin{align}
\mathcal{C}_{ij}\bar{\lambda}_i\bar{\lambda}_j
\end{align}
in the one-loop correction.

Note that the scale-derivative of the integral in the one-loop correction can also be obtained through a derivative with respect to the lower integration boundary. This does not require to evaluate the integral first, which is advantageous in certain situations. Indeed, we have
\begin{align}
 \frac{\mbox{d}}{\mbox{d}\ln b} \int_{\textbf{q}_\perp} \frac{1}{q_\perp^4} &= \frac{\mbox{d}}{\mbox{d}\ln b} \frac{1}{2\pi} \int_{\Omega/b}^\Omega \mbox{d}q_\perp \frac{1}{q_\perp^3}\\
 &=-\Bigl( \frac{\mbox{d}}{\mbox{d}\ln b} \frac{\Omega}{b}\Bigr)\frac{1}{2\pi} \frac{1}{(\Omega/b)^3}\\
 &=-\Bigl( - \frac{\Omega}{b}\Bigr)\frac{1}{2\pi} \frac{1}{(\Omega/b)^3}\\
 &=\frac{b^2}{2\pi \Omega^2}.
\end{align}
Consider then a more complicated model with one-loop correction
\begin{align}
 \delta \bar{\lambda} =  \bar{\lambda}(1)^2 \sum_{q_z}\int_{\textbf{q}_\perp}^\prime \frac{\mathcal{C} q_\perp^4+\mathcal{D}q_\perp^2( q_z^2+\alpha)+\mathcal{E}(q_z^2+\alpha)^2}{(q_\perp^2+q_z^2+\alpha)^4(\gamma q_\perp^2+q_z^2+\alpha)^2},
\end{align}
which models the case of confinement in the z-direction. Here $\gamma$ is a dimensionless parameter, $\alpha = -\pi^2/L_z^2$, $q_z=\pi n/L_z$, $n=1,2,\dots$, and 
\begin{align}
 \sum_{q_z}(\dots) = \frac{1}{L_z} \sum_{n=1}^\infty(\dots).
\end{align}
We have
\begin{align}
\nonumber  &\frac{\mbox{d}}{\mbox{d}\ln b} \int_{\textbf{q}_\perp}^\prime \frac{\mathcal{C} q_\perp^4+\mathcal{D}q_\perp^2(q_z^2+\alpha)+\mathcal{E} (q_z^2+\alpha)^2}{(q_\perp^2+q_z^2+\alpha)^2(\gamma q_\perp^2+q_z^2+\alpha)^2}\\
 \nonumber &= -\Bigl(-\frac{\Omega}{b}\Bigr)\frac{1}{2\pi} \frac{\mathcal{C} (\Omega/b)^4+\mathcal{D}(\Omega/b)^2(q_z^2+\alpha)+\mathcal{E} (q_z^2+\alpha)^2}{[(\Omega/b)^2+ q_z^2+\alpha]^2[\gamma (\Omega/b)^2+q_z^2+\alpha]^2}\\
 &=\frac{b^3}{2\pi \Omega^3} \frac{\mathcal{C}+\mathcal{D} \frac{b^2}{\Omega^2}(q_z^2+\alpha)+\mathcal{E}\frac{b^4}{\Omega^4}(q_z^2+\alpha)^2}{\gamma^2[1+\frac{b^2}{\Omega^2}(q_z^2+\alpha)]^2[1+\frac{1}{\gamma}\frac{b^2}{\Omega^2}(q_z^2+\alpha)]^2}.
\end{align}
The total one-loop correction can be written as
\begin{align}
 \frac{\mbox{d}\delta \bar{\lambda}}{\mbox{d}\ln b} = \bar{\lambda}(1)^2 \frac{b^2}{2\pi \Omega^2} h(\tilde{L}_z)
\end{align}
with 
\begin{align}
h(\tilde{L}_z) = \frac{1}{\tilde{L}_z}\sum_{n=1}^\infty \frac{\mathcal{C}+\mathcal{D} \frac{b^2}{\Omega^2}(q_z^2+\alpha)+\mathcal{E}\frac{b^4}{\Omega^4}(q_z^2+\alpha)^2}{\gamma^2[1+\frac{b^2}{\Omega^2}(q_z^2+\alpha)]^2[1+\frac{1}{\gamma}\frac{b^2}{\Omega^2}(q_z^2+\alpha)]^2}.
\end{align}
In this expression, the $L_z$-dependence is only due to the dimensionless combination $\tilde{L}_z=\Omega L_z/b$, because
\begin{align}
\frac{b^2}{\Omega^2}( q_z^2+\alpha) = \frac{\pi^2}{\tilde{L}_z^2}(n^2-1).
\end{align}
The flow equation for $\bar{\lambda}(b)$ becomes
\begin{align}
 \frac{\mbox{d}\bar{\lambda}}{\mbox{d}\ln b} = 2\bar{\lambda}+\frac{1}{2\pi \Omega^2} \bar{\lambda}^2 h(\tilde{L}_z).
\end{align}

\subsection{Flow of quartic couplings}\label{AppFlowQuart}

The one-loop corrections to the quartic couplings depend on  $q_z = \pi n/L_{\rm z}$, $n\geq 1$, only through the combination $q_z^2+\alpha$. By setting $\alpha=-q_z^2$ and neglecting the sum over $q_z$, we formally obtain the 2D limit. In the 2D limit, we have
\begin{align}
 \delta \beta_a =  -\frac{\mathcal{C}_a(\gamma)}{\gamma^2}\int_{\textbf{q}_\perp}^\prime\frac{1}{4q_\perp^4}
\end{align}
with
\begin{widetext}
\begin{align}
 \nonumber \mathcal{C}_1(\gamma) ={}&  24(\kappa^2+2\kappa+2)\beta_1^2+12(\kappa+2)^2\beta_1\beta_2+4(5\kappa^2+8\kappa+8)\beta_1\beta_3+8(\kappa^2+2\kappa+2)\beta_1\beta_4\\
&+24(\kappa^2+2\kappa+2)\beta_1\beta_5+4\kappa^2\beta_2\beta_5+2(5\kappa^2+16\kappa+16)\beta_3\beta_5+2\kappa^2\beta_4\beta_5+3\kappa^2\beta_5^2\\
\nonumber \mathcal{C}_2(\gamma) ={}& 4(\kappa^2+8\kappa+8)\beta_1^2+16(\kappa^2+2\kappa+2)\beta_1\beta_2+4\kappa^2\beta_1\beta_3+4\kappa^2\beta_1\beta_4+4\kappa^2\beta_1\beta_5+2(17\kappa^2+40\kappa+40)\beta_2^2\\
\nonumber &+4(9\kappa^2+16\kappa+16)\beta_2\beta_3+4(11\kappa^2+20\kappa+20)\beta_2\beta_4+24(\kappa^2+2\kappa+2)\beta_2\beta_5+2(3\kappa^2+4\kappa+4)\beta_3^2\\
\nonumber &+16(\kappa^2+\kappa+1)\beta_3\beta_4+2(5\kappa^2+8\kappa+8)\beta_3\beta_5+12(\kappa^2+2\kappa+2)\beta_4^2+2(5\kappa^2+8\kappa+8)\beta_4\beta_5\\
&+(\kappa^2+8\kappa+8)\beta_5^2\\
\nonumber \mathcal{C}_3(\gamma) ={}&4\kappa^2\beta_1^2+4\kappa^2\beta_1\beta_3+4(\kappa^2+8\kappa+8)\beta_1\beta_4+4(\kappa^2+8\kappa+8)\beta_1\beta_5+2\kappa^2\beta_2^2+12(\kappa+2)^2\beta_2\beta_3+4\kappa^2\beta_2\beta_4\\
&+2(7\kappa^2+8\kappa+8)\beta_3^2+16(\kappa^2+5\kappa+5)\beta_3\beta_4+2(\kappa^2+8\kappa+8)\beta_3\beta_5+4\kappa^2\beta_4^2+2\kappa^2\beta_4\beta_5+\kappa^2\beta_5^2\\
\nonumber \mathcal{C}_4(\gamma) ={}&4\kappa^2\beta_1^2+4(\kappa^2+8\kappa+8)\beta_1\beta_3+4\kappa^2\beta_1\beta_4+4(5\kappa^2+8\kappa+8)\beta_1\beta_5+2\kappa^2\beta_2^2+4\kappa^2\beta_2\beta_3+12(\kappa+2)^2\beta_2\beta_4\\
\nonumber &+2(3\kappa^2+20\kappa+20)\beta_3^2+16(\kappa^2+\kappa+1)\beta_3\beta_4+2\kappa^2\beta_3\beta_5+4(3\kappa^2+10\kappa+10)\beta_4^2+2(\kappa^2+8\kappa+8)\beta_4\beta_5\\
&+(17\kappa^2+32\kappa+32)\beta_5^2\\
\nonumber \mathcal{C}_5(\gamma) ={}& 8\kappa^2\beta_1\beta_2+8(\kappa+2)^2\beta_1\beta_3+16(\kappa^2+2\kappa+2)\beta_1\beta_4+16(\kappa^2+3\kappa+3)\beta_2\beta_5\\
&+4(3\kappa^2+4\kappa+4)\beta_3\beta_5+4(7\kappa^2+16\kappa+16)\beta_4\beta_5+6(\kappa+2)^2\beta_5^2.
\end{align}
\end{widetext}
Here we write $\kappa=\gamma-1$. For $\gamma=1$, these corrections agree with the result of Ref. \cite{Jones:1976zz} for the choice of symmetry group $\text{SO}(3)_{\rm S}\times \text{SO}(2)_{\rm L}\times\text{U}(1)$. 

In the quasi-2D case, we have
\begin{align}
 \alpha = - \frac{\pi^2}{L_z^2},
\end{align}
and so $q_z^2+\alpha = \Delta q_z^2$ with
\begin{align}
\Delta q_z^2 = \frac{\pi^2}{L_z^2}(n^2-1),\ n\geq 1.
\end{align}
The modes with $n>1$ are massive, but can lead to a quantitative effect in the RG equations. The flow equations in the quasi-2D regime have the form
\begin{align}
 \delta \bar{\beta}_a = -\sum_{q_z}\int_{\textbf{q}_\perp}^\prime \frac{\mathcal{C}_a(\gamma)q_\perp^4+\mathcal{C}_a(1)[(\gamma+1)q_\perp^2\Delta q_z^2+ \Delta q_z^4]}{4(q_\perp^2+\Delta q_z^2)^2(\gamma q_\perp^2+\Delta q_z^2)^2}.
\end{align}
The coefficients $\mathcal{C}_a(1)$ appear because the integrand needs to be a function of the three-dimensional rotation invariant $q_\perp^2+q_z^2+\alpha$ for $\gamma=1$. We have
\begin{align}
 \nonumber \mathcal{C}_1(1) ={}& 16(3\beta_1^2+3\beta_1\beta_2+2\beta_1\beta_3+\beta_1\beta_4\\
 &+3\beta_1\beta_5+2\beta_3\beta_5),\\
 \nonumber \mathcal{C}_2(1) ={}& 8(4\beta_1^2+4\beta_1\beta_2+10\beta_2^2+8\beta_2\beta_3+10\beta_2\beta_4\\
 \nonumber &+6\beta_2\beta_5+\beta_3^2+2\beta_3\beta_4+2\beta_3\beta_5+3\beta_4^2\\
 &+2\beta_4\beta_5+\beta_5^2),\\
 \nonumber \mathcal{C}_3(1) ={}& 16(2\beta_1\beta_4+2\beta_1\beta_5+3\beta_2\beta_3+\beta_3^2\\
 &+5\beta_3\beta_4+\beta_3\beta_5),\\
 \nonumber \mathcal{C}_4(1) ={}& 8(4\beta_1\beta_3+4\beta_1\beta_5+6\beta_2\beta_4+5\beta_3^2+2\beta_3\beta_4\\
 &+5\beta_4^2+2\beta_4\beta_5+4\beta_5^2),\\
\nonumber \mathcal{C}_5(1) ={}& 8(4\beta_1\beta_3+4\beta_1\beta_4+6\beta_2\beta_5+2\beta_3\beta_5\\
 &+8\beta_4\beta_5+3\beta_5^2).
\end{align}
Since the momentum components $q_z$ are not affected by the momentum shell cutoff, we can analytically evaluate the sum over the modes $q_z$. For $\gamma=1$, we have
 \begin{align}
 \delta \bar{\beta}_a|_{\gamma=1} = -\frac{\mathcal{C}_a(1)}{L_z}\sum_{n=1}^\infty\int_{\textbf{q}_\perp}^\prime \frac{1}{4(q_\perp^2+\Delta q_z^2)^2}.
\end{align}
For $\gamma\neq 1$, we have 
\begin{align}
 \frac{\mbox{d}\delta \bar{\beta}_a}{\mbox{d}\ln b} = -\frac{b^2}{8\pi \Omega^2} H_a(\gamma,\tilde{L}_z)
\end{align}
with 
\begin{align}
H_a(\gamma,\tilde{L}_z) = \frac{1}{\tilde{L}_z}\sum_{n=1}^\infty\frac{\mathcal{C}_a(\gamma)+\mathcal{C}_a(1)[(\gamma+1) \Delta \tilde{q}_z^2+\Delta \tilde{q}_z^4]}{\gamma^2[1+\Delta \tilde{q}_z^2]^2[1+\frac{1}{\gamma}\Delta \tilde{q}_z^2]^2}
\end{align}
and 
\begin{align}
 \Delta \tilde{q}_z^2 = \frac{\pi^2}{\tilde{L}_z^2}(n^2-1).
\end{align}
The summation over $n$ can be performed analytically.  The flow equations for the rescaled couplings 
\begin{align}
 \beta_a(b) = \frac{\bar{\beta}_a(b)}{8\pi \Omega^2}
\end{align}
read
\begin{align}
 \frac{\mbox{d}\beta_a}{\mbox{d}\ln b} = 2\beta_a - H_a(\gamma,\tilde{L}_z).
\end{align}
For $\gamma=1$ we have
\begin{align}
 H_a(1,\tilde{L}_z) &= \frac{\mathcal{C}_a(1)}{\tilde{L}_z}\sum_{n=1}^\infty \frac{1}{(1+\Delta\tilde{q}_z^2)^2}\\
 &=\mathcal{C}_a(1)\frac{\tilde{L}_z^3}{4}\frac{-2+\eta\ \coth(\eta)+\eta^2/\sinh^2(\eta)}{\eta^4}>0
\end{align}
with $\eta=\sqrt{\tilde{L}_z^2-\pi^2}$. Here we assumed $\eta>0$ as applies to the physical system.

\subsection{Flow of kinetic coefficient}

The flow equation for the parameter $\gamma$ can be obtained from Ref. \cite{Jones:1976zz}. To fix the normalization with respect to our choice of $\beta_a$, we denote the quartic coefficients in Ref. \cite{Jones:1976zz} as $\tilde{\beta}_a$. The flow of $\tilde{\beta}_1$ for $(m,n)=(3,2)$ in the 2D limit reads
\begin{align}
 \dot{\tilde{\beta}}_1 &= 2\tilde{\beta}_1 - \frac{1}{4\pi^2}mn\tilde{\beta}_1^2 +\dots\\
 &=2\tilde{\beta}_1 - \frac{1}{4\pi^2}6\tilde{\beta}_1^2 +\dots
\end{align}
Compared to our flow equation, in the 2D limit and for $\gamma=1$, given by
\begin{align}
 \dot{\beta}_1 &= 2 \beta_1 - 48 \beta_1^2+\dots,
\end{align}
we conclude that both schemes are related via the rescaling
\begin{align}
\frac{1}{4\pi^2}\tilde{\beta}_a = 8 \beta_a\ \Rightarrow\ \beta_a = \frac{1}{2}\times\frac{1}{16\pi^2}\tilde{\beta}_a.
\end{align}
The beta function for the kinetic coefficient $\gamma$ is given by
\begin{align}
 \dot{\gamma}=\gamma^2\frac{1}{(16\pi^2)^2}\frac{4}{3}\frac{1}{\gamma} \Bigl(\frac{1}{\gamma}-1\Bigr)\Bigl(\frac{1}{\gamma^2}+3\Bigr)\tilde{f}_\gamma\\
 =\frac{1}{(16\pi^2)^2}\frac{4}{3}(1-\gamma)\Bigl(\frac{1}{\gamma^2}+3\Bigr)\tilde{f}_\gamma,
\end{align}
where $\tilde{f}_\gamma>0$ is a positive definite quadratic form of the quartic couplings taken from Ref. \cite{Jones:1976zz}. Consequently, the prefactor $\frac{1}{(16\pi^2)^2}$ changes to $4$ when the expressed in terms of the rescaled couplings,
\begin{align}
 \frac{1}{(16\pi^2)^2} \tilde{f}_\gamma(\tilde{\beta}_a) = 4f_\gamma(\beta_a).
\end{align}
We then have
\begin{align}
 \dot{\gamma}=\frac{16}{3} (1-\gamma)\Bigl(\frac{1}{\gamma^2}+3\Bigr)f_\gamma
\end{align}
with
\begin{align}
 \nonumber f_\gamma={}& 12 \beta_1^2+2\beta_1\beta_2+8\beta_1\beta_3+2\beta_1\beta_4+6\beta_1\beta_5+\frac{13}{2}\beta_2^2\\
 \nonumber &+4\beta_2\beta_3+7\beta_2\beta_4+5\beta_2\beta_5+8\beta_3^2+4\beta_3\beta_4\\
 &+\frac{13}{2}\beta_4^2+5\beta_4\beta_5+\frac{15}{2}\beta_5^2.
\end{align}
The flow of $\gamma$ has an infrared-stable fixed pint at $\gamma_\star=1$. The stability is guaranteed due to $f_\gamma>0$ for any set of values of the quartic couplings. The initial value for the flow of $\gamma(b)$ is $\gamma(1)=3$ from mean-field theory.

\end{appendix}

\bibliography{refs_he3}


\end{document}